\documentclass[twocolumn,trackchanges]{aastex701}

\begin{document}

\title{\bf The non-equilibrium state of the Galactic disk $-$ new insights from GALAH DR4}

\author[orcid=0009-0005-7565-2817,gname=Peng,sname=Yang]{Peng Yang}
\affiliation{Local Universe and Time-Domain Astronomy Laboratory, LuoxiaHong Institute of Astronomy, China West Normal University, Nanchong 637002, China}
\email{yangpengastro@gmail.com}

\author[orcid=0000-0001-8459-1036, gname=Hai-Feng,sname=Wang]{Hai-Feng Wang}
\affiliation{Local Universe and Time-Domain Astronomy Laboratory, LuoxiaHong Institute of Astronomy, China West Normal University, Nanchong 637002, China}
\email[show]{haifeng.wang.astro@gmail.com} 

\author[orcid=0000-0001-7516-4016,gname=Joss,sname=Bland-Hawthorn]{Joss Bland-Hawthorn}
\affiliation{Sydney Institute for Astronomy, School of Physics A28, University of Sydney, Camperdown, NSW 2006, Australia}
\email{jonathan.bland-hawthorn@sydney.edu.au}

\author[orcid=0009-0001-8495-4164, gname=Guo-Jun,sname=Cheng]{Guo-Jun Cheng}
\affiliation{Local Universe and Time-Domain Astronomy Laboratory, LuoxiaHong Institute of Astronomy, China West Normal University, Nanchong 637002, China}
\email{chengguojunastro@gmail.com} 

\author[orcid=0009-0005-1184-2224, gname=Yu-Zheng,sname=Wang]{Yu-Zheng Wang}
\affiliation{Department of Physics, Hebei Normal University, Shijiazhuang, 050024, China}
\email{swordzun@gmail.com} 

\author[orcid=0000-0003-3736-6076, gname=Yangping,sname=Luo]{Yang-Ping Luo}
\affiliation{Department of Astronomy, China West Normal University, Nanchong, 637002, China}
\email{luoyangping789@163.com} 

\author[orcid=0000-0002-5745-827X, gname=Kun,sname=Wang]{Kun Wang}
\affiliation{Department of Astronomy, China West Normal University, Nanchong, 637002, China}
\email{kwang@cwnu.edu.cn} 

\begin{abstract}
The non-equilibrium state of the Galactic disk has drawn extensive attention since ESA's \textit{Gaia} launch. Complex structure appears across a multi-parameter space of stellar abundances, ages, and kinematics. Using GALAH DR4, we investigate ridge and phase-spiral structures in chemical-abundance, phase ($v_i,x_i$), action--angle ($J_i,\theta_i$), and frequency--action ($\Omega_i,J_i$) spaces ($i=R,\phi,z$). Our sample includes 333,412 disk stars with reliable chemodynamical data. In the $R$--$v_\phi$ plane, ridges trace near-constant $J_\phi$ tracks and show coherent variations in $\theta_R$ and $\theta_\phi$. In $J_\phi$--$J_R$, ridges are most prominent when encoded by $z_{\rm max}$ and preferentially associate with metal-rich, low-$\alpha$ populations. In the solar neighbourhood ($R=7.2$--$9.2$\,kpc), we recover a clear phase spiral in $z$--$v_z$. Encoded in $v_\phi$, $\Omega_R$, or $\Omega_\phi$, one-armed spirals appear, whereas the $v_R$-encoded map shows a two-armed spiral. Projecting actions onto $z$--$v_z$ reveals residual spiral signatures in $J_R$ and $J_\phi$, indicating that the phase spiral encodes in-plane orbital information. Both structures depend on age: ridges remain at similar locations with varying contrast, while the phase spiral is more tightly wound in younger and more loosely wound in older populations. Several ridges correspond in $J_\phi$ to known moving groups, and excluding stars in all six moving-group regions makes the phase spiral less distinct in mean radial velocity maps. These links support interpreting ridges, phase spirals, and moving groups within a common phase-space dynamical framework. Combined with trends in orbital properties, age, and abundance, these findings suggest a common dynamical origin and distinct chemodynamical responses of a perturbed Galactic disk.
\end{abstract}



\section{Introduction}
\label{sec:introduction}

The Milky Way provides an unparalleled laboratory for investigating the dynamical evolution of disk galaxies through the phase-space, chemical, and age distributions of individual stars \citep{Freeman2002,Rix2013,BlandHawthorn2016,Helmi2020}. Although smooth, axisymmetric, and approximately steady-state models provide a useful baseline description of the Galactic potential and stellar orbits \citep{Binney2008,Bovy2015,McMillan2017}, observations from \textit{Gaia} have revealed prominent phase-space substructures throughout the disk \citep{GaiaCollaboration2018,Antoja2018,GaiaCollaboration2023,Hunt2025}. These features demonstrate substantial departures from dynamical equilibrium and preserve information about the response of the disk to internal non-axisymmetric structures and external perturbations.

Among the most prominent in-plane features are the ridges observed as inclined overdensities in the $R$--$v_\phi$ plane (Galactic radius and azimuthal velocity) \citep{Kawata2018,Ramos2018,Khanna2019,Wang2020,Yang2023}. The ridges are aligned with curves of constant $L_z$ and related substructures have also been identified in action space or integrals-of-motion space, suggesting that they trace coherent orbital structure rather than arising solely from fluctuations in the projected number density \citep{Trick2019,GaiaCollaboration2023}. Their physical origin remains debated, with interpretations invoking either resonances with the Galactic bar or spiral arms, transient spiral structure, radial migration, or phase mixing following a dynamical perturbation \citep{Hunt2018,Trick2019,Laporte2019,Li2024}.

The phase spiral provides a complementary probe of disequilibrium in the Galactic disk. First identified in the $z$--$v_z$ plane (vertical distance and velocity) among stars in the solar neighborhood \citep{Antoja2018}, it has since been detected over Galactocentric radii of $6$--$11\,\mathrm{kpc}$, with its morphology varying with angular momentum and azimuth \citep{Antoja2023}. It is commonly interpreted as a signature of incomplete vertical phase mixing: following a perturbation, stars with different vertical oscillation frequencies accumulate phase differences, gradually winding the perturbed stellar distribution into a spiral \citep{Antoja2018,Binney2018,BlandHawthorn2021,Antoja2023}. Modelling the phase spiral in vertical frequency--angle space has revealed variations in its amplitude and inferred winding time with angular momentum, indicating that a single perturbation time does not adequately describe all orbital subsamples \citep{Frankel2023}. However, the complex morphology of the spiral and its dependence on orbital properties suggest that its present-day structure may reflect multiple perturbations, self-gravitating disk responses, or sustained forcing.

Ridges and phase spirals are often regarded as signatures of stellar motion in different directions: ridges primarily trace in-plane dynamics, whereas phase spirals characterize vertical disequilibrium. Nevertheless, the radial and vertical degrees of freedom of disk-star orbits are unlikely to evolve independently. A perturbation may simultaneously excite in-plane and vertical motions, while the subsequent response can be modified by the bar, spiral structure, and the self-gravity of the disk. 

Observational links among ridges, moving groups, vertical waves, and the action dependence of the phase spiral therefore raise the possibility that these structures encode different aspects of a coupled three-dimensional response \citep{Khanna2019,Laporte2019,BlandHawthorn2021,Hunt2022}. Whether the observed correlations reflect a common perturbation, the interaction of several mechanisms, or the subsequent coupling of initially distinct structures remains unsolved.

Both external and internal mechanisms can contribute to this nonequilibrium state. Interactions with the Sagittarius dwarf galaxy can excite bending waves, vertical oscillations, phase spirals, and large-scale in-plane responses \citep{Gomez2013,Laporte2018,Laporte2019,BlandHawthorn2021}. Recent simulations show that a significant Sagittarius-like disk crossing approximately $0.7$--$1.2\,\mathrm{Gyr}$ ago can reproduce several broad properties of the observed disk corrugations and phase spiral, although it does not simultaneously reproduce all in-plane wave-like structures in detail \citep{Yamsiri2026}. Internal non-axisymmetric processes, including the Galactic bar, spiral arms, resonances, transient spirals, and radial migration, can generate moving groups and ridges in the $R$--$v_\phi$ and $L_z$--$J_R$ (angular momentum and radial action) planes \citep{Dehnen2000,SellwoodBinney2002,MinchevFamaey2010,Hunt2018,Khanna2019,Trick2019,Khoperskov2020}, and bar evolution may also excite vertical oscillations and phase-space spirals \citep{Khoperskov2019,Chiba2021}.

Gravitational perturbations from the Large Magellanic Cloud, including those mediated by the wake it induces in the Milky Way's dark matter halo, may affect the vertical structure of the Galactic disk, most notably the retrograde precession of the outer disc warp \citep{Laporte2018,GaravitoCamargo2019}. Dark matter subhalos can also excite bending waves \citep{Chequers2018} and perturb the vertical phase-space distribution of disk stars \citep{Gilman2025}. In addition, isolated $N$-body+SPH simulations indicate that irregular gas accretion along a warped disk can continuously excite bending waves and generate widespread, long-lived phase spirals \citep{Khachaturyants2022, Wang2026}, or excited either by dark matter substructure or by a clumpy, turbulent interstellar medium \citep{Tremaine2023,TepperGarcia2025}. These mechanisms are not mutually exclusive and may operate simultaneously or over different timescales.

Despite this progress, joint observational studies of ridges and phase spirals using a homogeneous stellar sample, a common dynamical model, and simultaneous age and chemical information remain limited. In particular, it is not yet clear how the morphology and kinematic contrast of the phase spiral depend jointly on the in-plane actions $J_R$ and $L_z$, whether these dependencies can be associated with ridge-like orbital structures, or whether ridges exhibit coherent vertical signatures that persist across different dynamical spaces. The age and chemical dependence of such correlations can further distinguish long-lived orbital structures from transient responses and can constrain whether the two phenomena are consistent with a common three-dimensional disk response.

In this work, we use GALAH DR4 to investigate ridge structures and phase spirals within a unified chemodynamical framework. We characterize their distributions in phase space, action space, frequency space, angle space, and integrals-of-motion space, focusing on the $R$--$v_\phi$, $L_z$--$J_R$, $L_z$--$E$, $z$--$v_z$, and $\Omega_z$--$\sqrt{J_z}$ planes. We examine the corresponding variations in stellar density, velocity, orbital phase, age, metallicity, and $\alpha$-element abundance. Our primary goals are to determine whether phase-spiral signatures vary systematically with in-plane orbital properties, whether ridge-associated structures contain coherent vertical-kinematic signatures, and how these correlations depend on stellar age and chemistry. These measurements provide empirical constraints on the coupling between in-plane and vertical motions and on the range of mechanisms that may contribute to the present nonequilibrium state of the Galactic disk.

This paper is organized as follows. In Section~\ref{sec:data_methods}, we describe the GALAH DR4 sample selection, the adopted chemical-abundance measurements, and the calculation of dynamical quantities in the Galactic potential of \citet{McMillan2017}. In Section~\ref{sec:results}, we present the ridge and phase-spiral structures identified in the different dynamical spaces and examine their age and chemical dependence. In Section~\ref{sec:discussion}, we discuss the possible connection between the two classes of substructure, their physical origins, and their implications for the nonequilibrium evolution of the Galactic disk. We summarize our main conclusions in Section~\ref{sec:conclusion}.


\newcommand{\feh}{[\mathrm{Fe/H}]}
\newcommand{\afe}{[\alpha/\mathrm{Fe}]}

\section{Data and methods}
\label{sec:data_methods}

\subsection{GALAH DR4 data and sample selection}
\label{subsec:galah_sample}

In this work, we use the fourth data release of the Galactic Archaeology with HERMES survey (GALAH DR4) to construct a chemodynamical sample of Milky Way disk stars. GALAH is a large high-resolution spectroscopic survey designed to investigate the formation and evolution of the Milky Way through detailed stellar parameters and chemical abundances \citep{DeSilva2015, Martell2017, Buder2025}. The survey is based on spectra obtained with the HERMES spectrograph on the Anglo-Australian Telescope \citep{Sheinis2015}. GALAH DR4 combines spectroscopic information with Gaia astrometry and provides stellar atmospheric parameters, elemental abundances, stellar ages, kinematics, and associated quality indicators \citep{Buder2025, GaiaDR3Summary2023, Lindegren2021}.

We adopt the Galactocentric cylindrical coordinates $(R,\phi,z)$ and velocity components $(v_R,v_\phi,v_z)$ provided by the GALAH DR4 dynamical catalogue \citep{Buder2025}. The adopted Galactocentric transformation assumes a Solar Galactocentric distance of $R_\odot=8.21\,{\rm kpc}$, a Solar height of $z_\odot=25\,{\rm pc}$, and a circular velocity of $V_{\rm circ}=233.1\,{\rm km\,s^{-1}}$ \citep{McMillan2017}. The Solar peculiar motion is $(U_\odot,V_\odot,W_\odot)=(11.1,15.17,7.25)\,{\rm km\,s^{-1}}$. We further use the GALAH DR4 metallicity $\feh$, stellar ages, Gaia astrometric-quality indicators, signal-to-noise ratios, and the relevant spectroscopic quality flags. Stellar ages are adopted from the GALAH DR4 estimates based on Bayesian stellar-parameter inference with BSTEP \citep{Sharma2018, Buder2025}.

To construct a reliable disk sample, we first require finite measurements for all quantities used in the analysis. We then restrict the sample to
\begin{equation}
6<R<12\,{\rm kpc}, \qquad |z|<1\,{\rm kpc}, \qquad 0<{\rm age}<14\,{\rm Gyr}.
\label{eq:spatial_age_cut}
\end{equation}
These limits are adopted to focus on the Galactic disk over the radial range relevant to our analysis while reducing contamination from stars at large vertical distances and excluding unphysical or poorly constrained stellar ages.

For the Gaia astrometric quality selection, we require a positive parallax and parallax uncertainty, together with
\begin{equation}
\varpi/\sigma_\varpi>5, \qquad {\rm RUWE}<1.4,
\label{eq:gaia_quality_cut}
\end{equation}
where $\varpi$ and $\sigma_\varpi$ denote the parallax and its uncertainty, respectively. The parallax signal-to-noise criterion corresponds to a relative parallax uncertainty below approximately 20\%, while the RUWE criterion removes sources with potentially unreliable astrometric solutions. For the GALAH spectroscopic measurements, we require \texttt{flag\_sp}=0, \texttt{flag\_sp\_fit}=0, \texttt{flag\_red}=0, and \texttt{flag\_fe\_h}=0, together with \texttt{snr\_px\_ccd3}$>30$.

We use the GALAH DR4 $\feh$ measurements as the stellar metallicity indicator and construct an $\alpha$-element abundance estimate from Mg, Si, Ca, and Ti. For each star, we define
\begin{equation}
\afe=\frac{1}{N_\alpha}\sum_X[X/{\rm Fe}], \qquad X={\rm Mg,Si,Ca,Ti},
\label{eq:alpha_fe}
\end{equation}
where $N_\alpha$ is the number of reliable elemental-abundance measurements available for a given star. Each elemental abundance is included only when its corresponding GALAH abundance quality flag is zero and \texttt{snr\_px\_ccd3}$>30$. We require at least three reliable measurements among Mg, Si, Ca, and Ti to determine $\afe$, thereby reducing the influence of uncertain measurements from individual elements. After applying all spatial, astrometric, spectroscopic, and chemical-abundance selection criteria, the final sample contains 333,412 stars.

\begin{figure*}[t]
\centering
\includegraphics[width=0.95\textwidth]{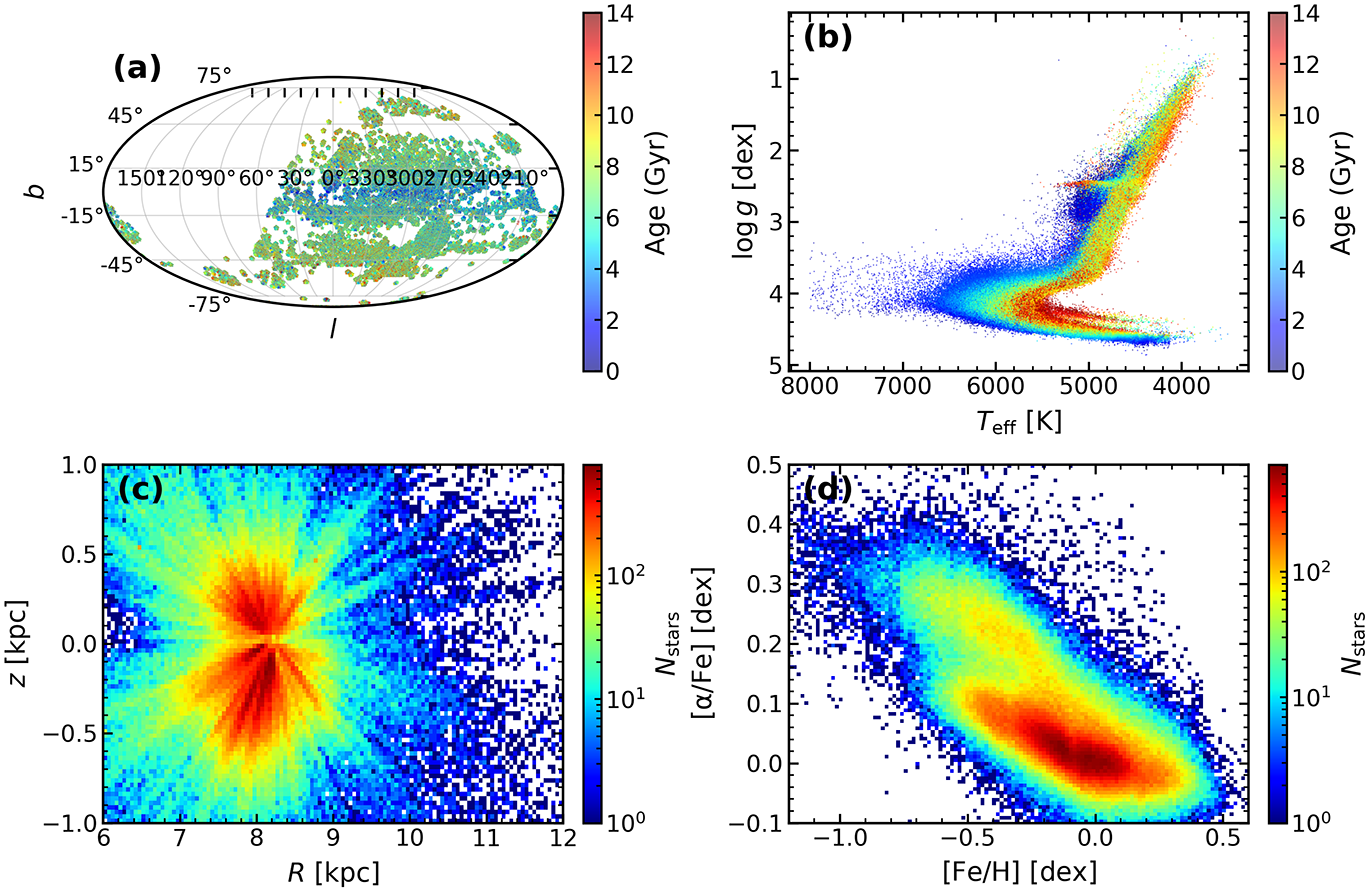}
\caption{
Overview of the final GALAH DR4 disk sample adopted in this work. The panels show the distributions of the sample in Galactic coordinates $(l,b)$, colour-coded by stellar age (a); in the $T_{\rm eff}$--$\log g$ plane, colour-coded by stellar age (b); in the Galactocentric $R$--$z$ plane, colour-coded by stellar number density (c); and in the $\feh$--$\afe$ plane, colour-coded by stellar number density (d).
}
\label{fig:galah_data_overview}
\end{figure*}

Fig.~\ref{fig:galah_data_overview} summarizes the basic properties of the final GALAH DR4 disk sample. The sky distribution reflects the observational footprint of GALAH, while the $T_{\rm eff}$--$\log g$ diagram shows that the sample contains stars spanning different evolutionary stages. The $R$--$z$ distribution illustrates the spatial coverage of the adopted disk sample, and the $\feh$--$\afe$ distribution shows the broad chemical-abundance coverage of the selected stars, providing the basis for the chemodynamical analysis presented below.

\subsection{Dynamical framework and orbital quantities}
\label{subsec:dynamical_quantities}

To investigate the orbital properties of the selected stars in a homogeneous dynamical framework, we recalculate the orbital quantities in the axisymmetric Milky Way potential of \citet{McMillan2017}, implemented in \texttt{galpy} \citep{Bovy2015}. We adopt the physical length and velocity scales $R_0=8.21\,{\rm kpc}$ and $V_0=233.1\,{\rm km\,s^{-1}}$, consistent with the scaling adopted for the GALAH DR4 dynamical catalogue. We note that the key conclusions are robust against the choice of solar motion or \texttt{AGAMA} calculation \citep{Vasiliev2019}.

\begin{figure*}[p]
\centering
\includegraphics[height=0.85\textheight]{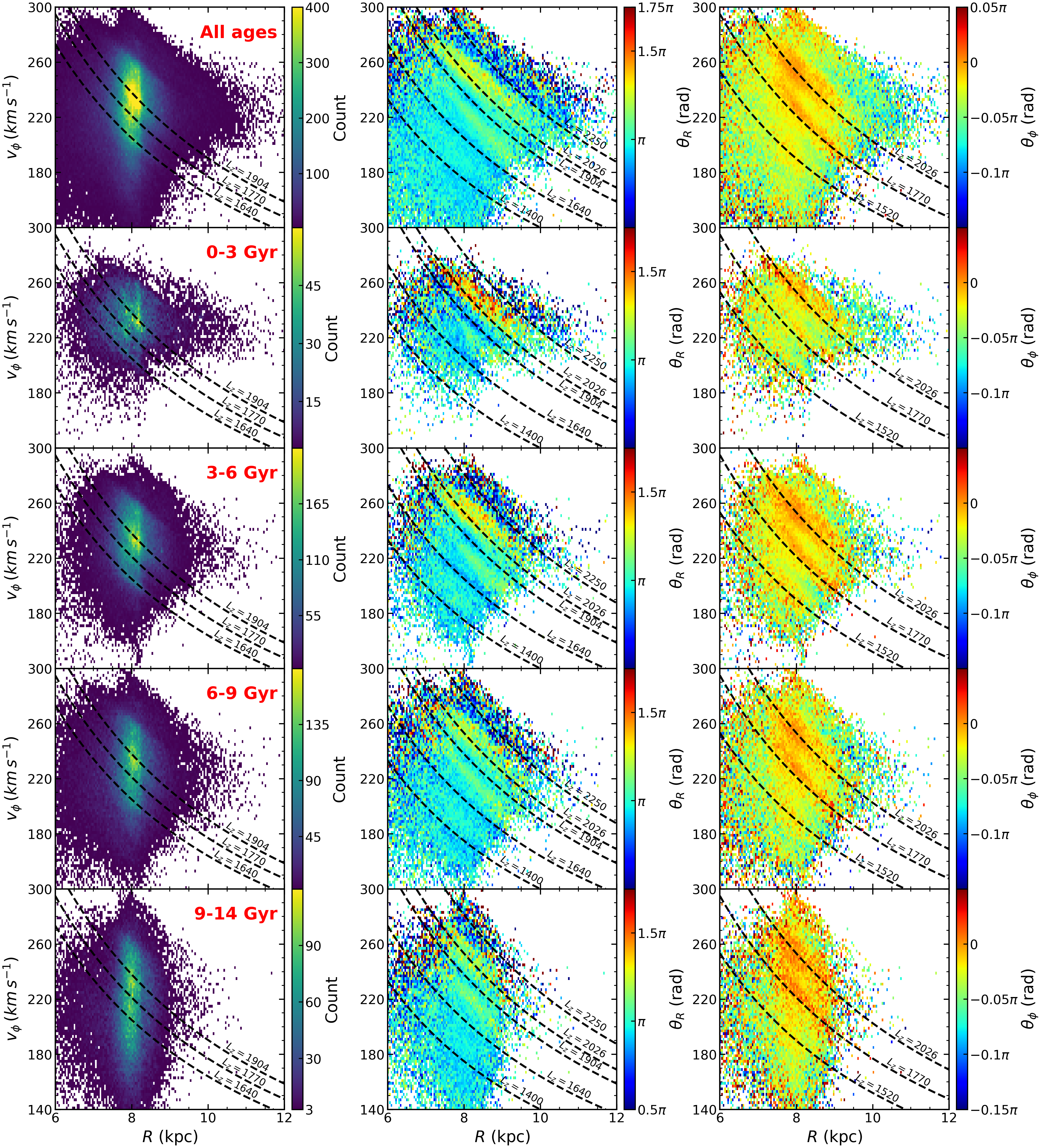}

\caption{\small Distribution of the GALAH DR4 disk sample in the $R$--$v_\phi$ plane for different stellar-age bins. From top to bottom, the rows correspond to the full sample and to stars with ages of $0$--$3$, $3$--$6$, $6$--$9$, and $9$--$14\,{\rm Gyr}$. The left, middle, and right columns show stellar counts, the mean radial angle $\theta_R$, and the mean azimuthal angle $\theta_\phi$, respectively. Representative curves of constant vertical angular momentum, $L_z=Rv_\phi$, are shown as black dashed lines. The ridge morphology is schematically illustrated in Fig.~\ref{fig:GALAH_Same_Dynamical_Framework}.}

\label{fig:GALAH_R_vphi_age}
\end{figure*}

We use the \texttt{actionAngleStaeckel} implementation in \texttt{galpy} to calculate the orbital actions, frequencies, and angle variables. This method is based on the St\"ackel-fudge approximation and provides an efficient estimate of orbital quantities in axisymmetric potentials \citep{Binney2012, SandersBinney2016}. We adopt a St\"ackel focal-length parameter of $\Delta=0.45$. For each star, we calculate the radial action $J_R$, azimuthal action $L_z$, and vertical action $J_z$, together with the corresponding orbital frequencies $(\Omega_R,\Omega_\phi,\Omega_z)$ and angle variables $(\theta_R,\theta_\phi,\theta_z)$. The calculation follows the same procedure as that adopted in the GALAH DR4 stellar-dynamics catalogue, but all orbital quantities are independently recalculated in this work. Broadly speaking, the radial action $J_R$ characterizes radial orbital excursions and reflects the degree of radial orbital excitation; the azimuthal action $J_\phi=L_z$ represents angular momentum about the Galactic symmetry axis and is closely related to the guiding-centre radius; and the vertical action $J_z$ quantifies vertical orbital oscillations and serves as an indicator of vertical orbital excitation and heating.

We also calculate the specific orbital energy as
\begin{equation}
E=\frac{1}{2}\left(v_R^2+v_\phi^2+v_z^2\right)+\Phi(R,z),
\label{eq:energy}
\end{equation}
where $\Phi(R,z)$ is the gravitational potential of the McMillan17 model. Although an axisymmetric potential is adopted for these calculations, this does not imply that the Galactic disk is assumed to be in complete dynamical equilibrium. Instead, the McMillan17 potential provides a smooth reference description of the mean Galactic gravitational field, allowing coherent non-equilibrium structures to be characterised in phase space, action space, frequency space, angle space, and integrals-of-motion space.

\begin{figure*}[htp!]
\centering
\includegraphics[width=0.95\textwidth]{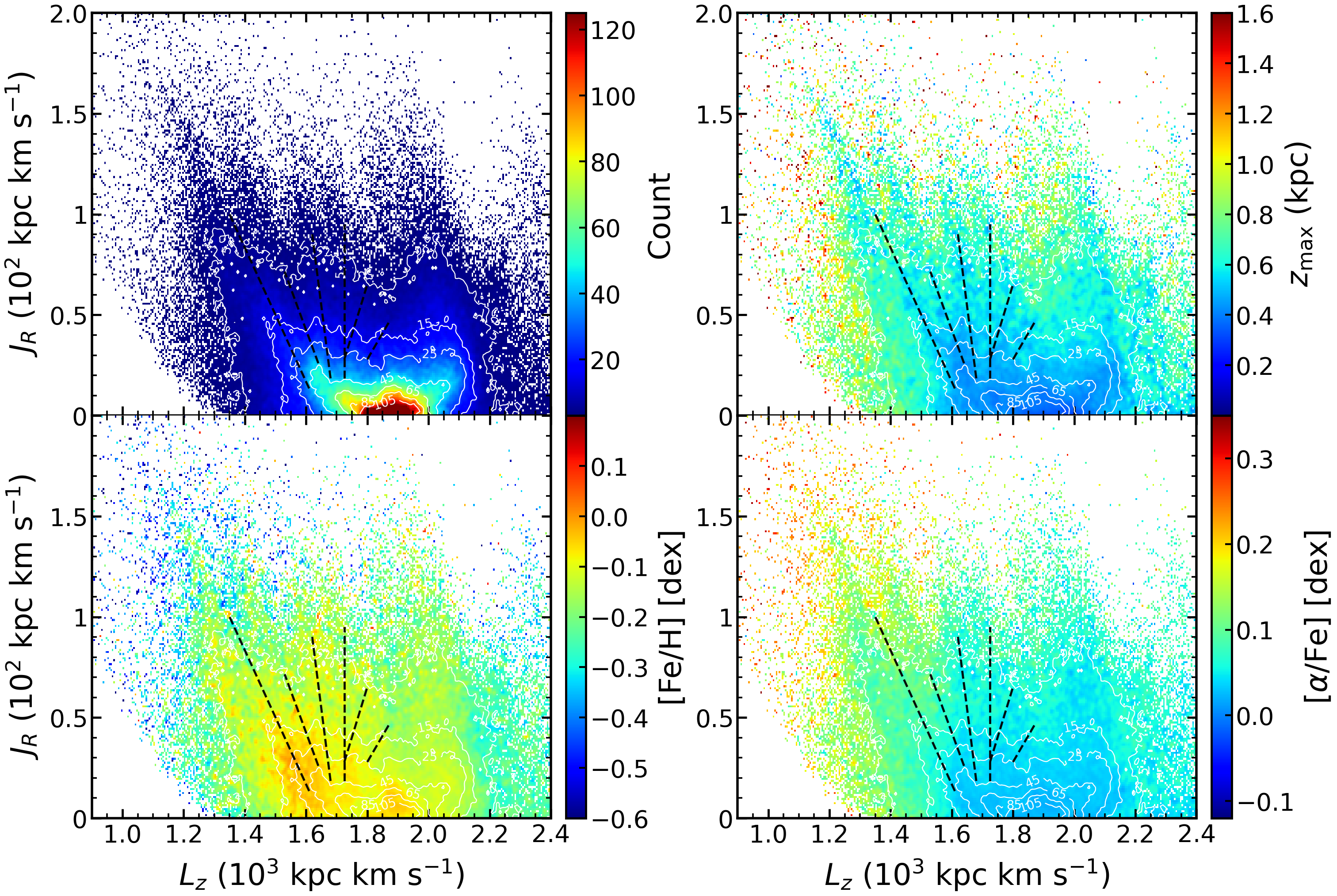}
\caption{\small Distribution of the GALAH DR4 disk sample in the $L_z$--$J_R$ action plane. The upper-left, upper-right, lower-left, and lower-right panels show stellar counts, maximum orbital height $z_{\rm max}$, metallicity $\feh$, and $\alpha$-element abundance $\afe$, respectively. The white contours trace the stellar number-density distribution, while the black dashed lines indicate the ridge-like structures identified in this plane. 
}
\label{fig:GALAH_Lz_Jr}
\end{figure*}

\section{Results}
\label{sec:results}

In this section, we present the non-equilibrium structures identified in the GALAH DR4 disk sample, focusing on ridges and phase spirals. We first characterize ridge structures in the $R$--$v_\phi$ plane and action-related spaces, including their associated orbital-angle signatures (Section~\ref{subsec:ridge_structures}). We then examine phase spirals in the vertical $z$--$v_z$ plane and their signatures in action-related spaces (Section~\ref{subsec:phase_spiral}). For both types of structure, we investigate how their morphology and associated orbital properties vary with stellar age and chemical abundance.

\subsection{Ridge structures in phase space and action-related spaces}
\label{subsec:ridge_structures}

To examine how the morphology of the ridges and their associated orbital-angle signatures vary with stellar age, Fig.~\ref{fig:GALAH_R_vphi_age} presents the distribution of the GALAH DR4 disk sample in the $R$--$v_\phi$ plane, comparing the full sample with four stellar-age bins.

In the stellar-count map of the full sample, several inclined ridges are clearly visible and approximately follow the constant-angular-momentum curves. Similar features are also visible at corresponding locations in the radial and azimuthal orbital-angle maps of $\theta_R$ and $\theta_\phi$. The age-separated maps show that several ridges remain identifiable at broadly similar positions and follow the same constant-angular-momentum curves across multiple age bins. However, their contrast in the $\theta_R$ and $\theta_\phi$ maps varies with stellar age. In the $\theta_R$ maps, the ridges following the $L_z=2026$ and $L_z=1904$ (km $s^{-1}$ kpc) curves exhibit stronger contrast in the $0$--$3$ and $3$--$6$~Gyr, but are less distinct in the $6$--$9$ and $9$--$14$~Gyr. In the $\theta_\phi$ maps, the ridges along the $L_z=2026$ and $L_z=1770$ curves show weaker contrast in the $0$--$3$~Gyr bin and stronger contrast in the other three age bins.

The presence of these ridges at similar angular momenta across multiple age bins suggests that different stellar-age populations may trace common dynamical structures in the disk. The more prominent $\theta_R$ signatures in the younger populations are consistent with the expectation that dynamically colder populations more readily exhibit perturbation-induced kinematic substructure. However, the stronger contrast of the $\theta_\phi$ signatures in the older bins suggests that these age-dependent differences may also be related to the orbital distributions of the populations and how these distributions are reflected in the different orbital-angle maps.

To complement the ridge analysis in the $R$--$v_\phi$ plane, we next examine the vertical orbital properties and chemical characteristics of structures in action space. Fig.~\ref{fig:GALAH_Lz_Jr} presents the distribution of the full sample in the $L_z$--$J_R$ plane, with the upper-left, upper-right, lower-left, and lower-right panels showing stellar counts, maximum orbital height $z_{\rm max}$, metallicity $\feh$, and $\alpha$-element abundance $\afe$, respectively. Several ridge-like sequences are more clearly delineated in the $z_{\rm max}$ map than in the stellar-count map. They extend from the low-$J_R$ region toward larger radial actions, with their locations broadly corresponding to local variations in stellar counts. Across several of these ridges, the $z_{\rm max}$ map reveals systematic changes in vertical orbital extent, indicating that the ridge structure is associated with differences in the vertical orbital properties of the different stellar populations.

The abundance maps provide a complementary view of these structures against a background of large-scale chemodynamical gradients. The low-$J_R$, intermediate-$L_z$ region is relatively metal-rich and $\alpha$-poor, whereas the lower-$L_z$, higher-$J_R$ region is generally more metal-poor and $\alpha$-enhanced. Comparing the abundance maps with the prominent ridges identified in the $z_{\rm max}$ map reveals that the ridge regions are generally associated with relatively metal-rich and low-$\alpha$ stellar populations.

\begin{figure*}[p]
\centering
\includegraphics[height=0.85\textheight]{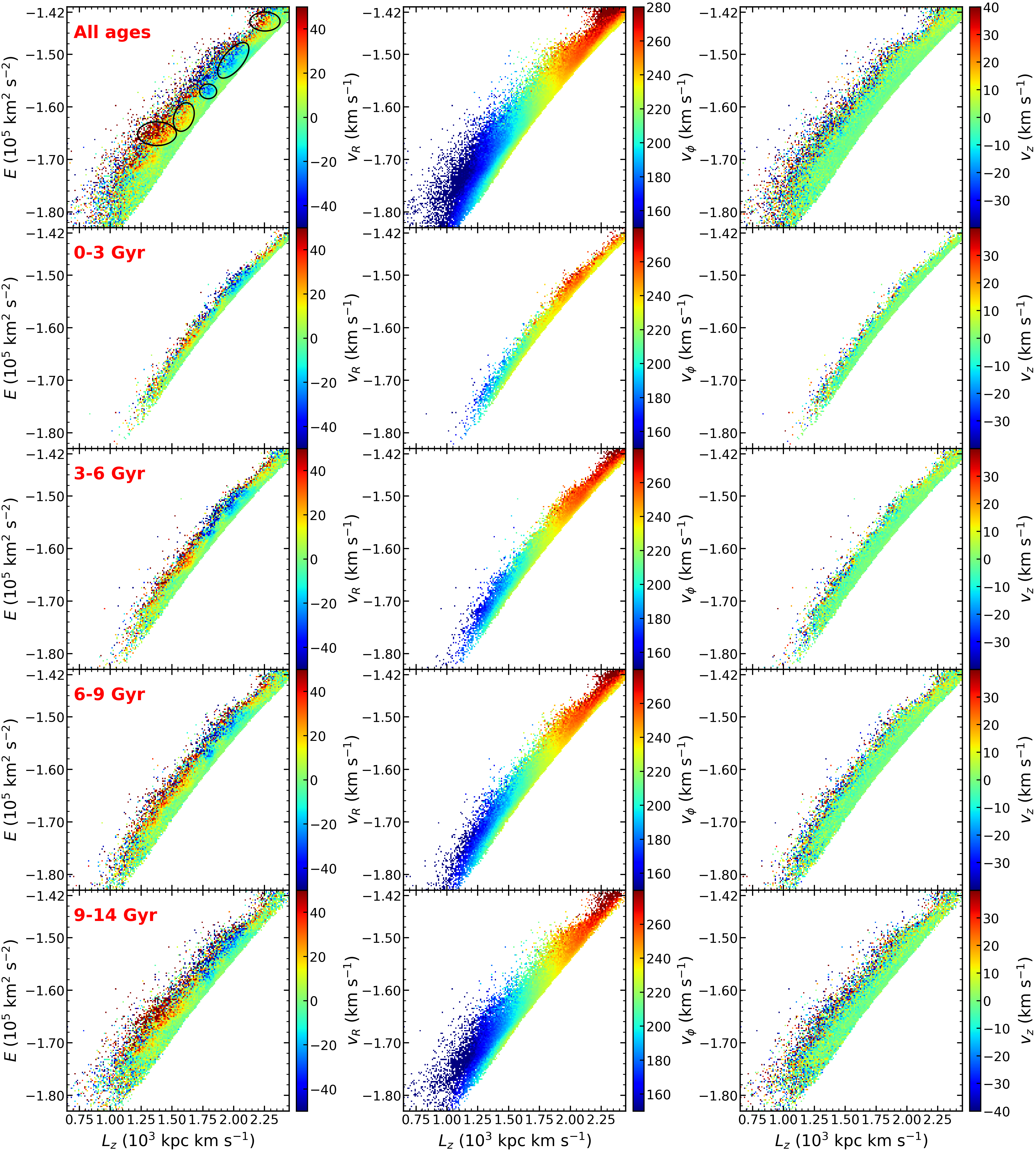}

\caption{\small Distribution of the GALAH DR4 disk sample in the $L_z$--$E$ plane for different stellar-age bins. From top to bottom, the rows correspond to the full sample and to stars with ages of $0$--$3$, $3$--$6$, $6$--$9$, and $9$--$14\,{\rm Gyr}$. From left to right, the columns are colour-coded by the mean Galactocentric velocity components $v_R$, $v_\phi$, and $v_z$, respectively. Black circles in the top-left panel highlight detected kinematic substructures visible in the mean $v_R$ map.}

\label{fig:GALAH_Lz_E_v}
\end{figure*}

\begin{figure*}[p]
\centering
\includegraphics[height=0.85\textheight]{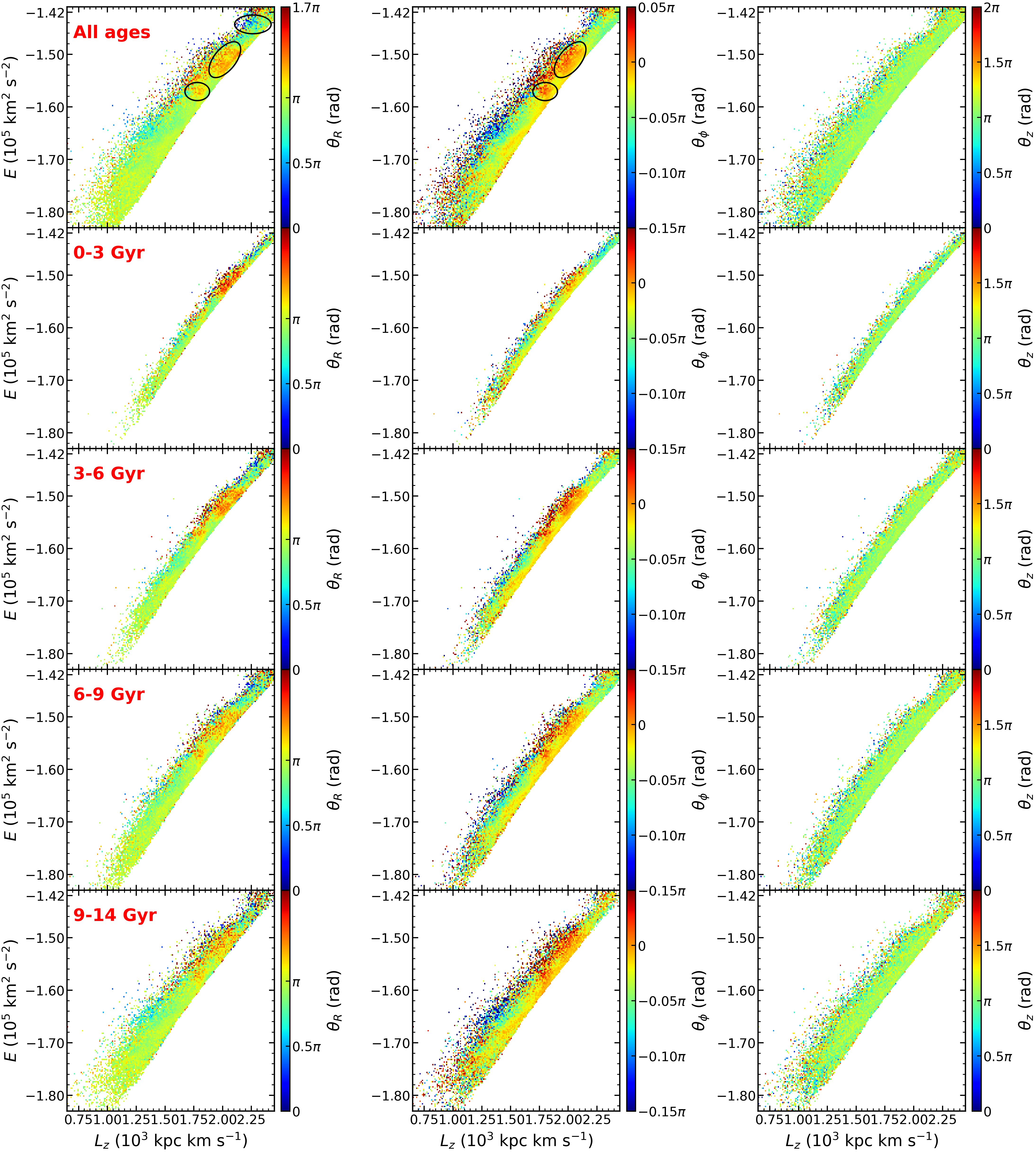}

\caption{\small Same as Fig.~\ref{fig:GALAH_Lz_E_v}, but colour-coded by the orbital angle variables. From left to right, the columns show the mean radial, azimuthal, and vertical angles, $\theta_R$, $\theta_\phi$, and $\theta_z$, respectively. Black circles in the top-left and top-middle panels highlight detected substructures in the mean $\theta_R$ and $\theta_\phi$ maps.}

\label{fig:GALAH_Lz_E_Theta}
\end{figure*}

Building on the analysis of the $R$--$v_\phi$ and $L_z$--$J_R$ planes, we further examine the distribution of ridge-related structures in angular momentum--energy space and how their velocity and orbital-phase signatures vary with stellar age. Figures~\ref{fig:GALAH_Lz_E_v} and \ref{fig:GALAH_Lz_E_Theta} present the velocity-component and orbital-angle maps in the $L_z$--$E$ plane, respectively. With increasing age, the occupied region broadens and extends toward lower $L_z$ and lower $E$ under the adopted energy convention, indicating that the older populations span a wider range of angular momentum and orbital energy.

The $v_R$ maps exhibit localised structures with alternating positive and negative radial velocities. Some of these features remain identifiable across multiple age intervals. By comparison, the $v_\phi$ maps are dominated by a large-scale trend with $L_z$, whereas no obvious local substructure or systematic trend is apparent in the $v_z$ maps. Among the velocity components shown, radial velocity therefore provides a relatively clear view of these local features.

The orbital-angle maps provide complementary phase information. Both the $\theta_R$ and $\theta_\phi$ maps exhibit the same local features, whereas no clear coherent substructure is apparent in $\theta_z$. The contrast of the local structures in the $\theta_R$ and $\theta_\phi$ maps varies with stellar age: the $\theta_R$ signatures show lower contrast in older stellar populations, while the $\theta_\phi$ signatures exhibit the opposite trend. These trends indicate that the visibility of the radial and azimuthal orbital-angle signatures depends on stellar age and highlight the additional information provided by orbital angles when comparing the dynamical properties of different stellar-age populations, implying that the oscillation amplitude of the angles, along with the heating history, evolved with time.

Taken together, the $R$--$v_\phi$, $L_z$--$J_R$, and $L_z$--$E$ distributions reveal ridge-related substructure in several dynamical representations, associated with variations in vertical orbital extent, in-plane velocity, and radial and azimuthal orbital angles. These correspondences connect ridge morphology with the orbital properties and phase distributions of the stars. The age-dependent contrast provides a basis for comparing ridge-related structures across stellar populations, while interpreting their origins requires considering the overall orbital distributions and chemical-abundance properties of the different age groups.

\subsection{Phase spirals in vertical phase space and action-related spaces}
\label{subsec:phase_spiral}

\begin{figure*}[!t]
\centering
\includegraphics[width=0.95\textwidth]{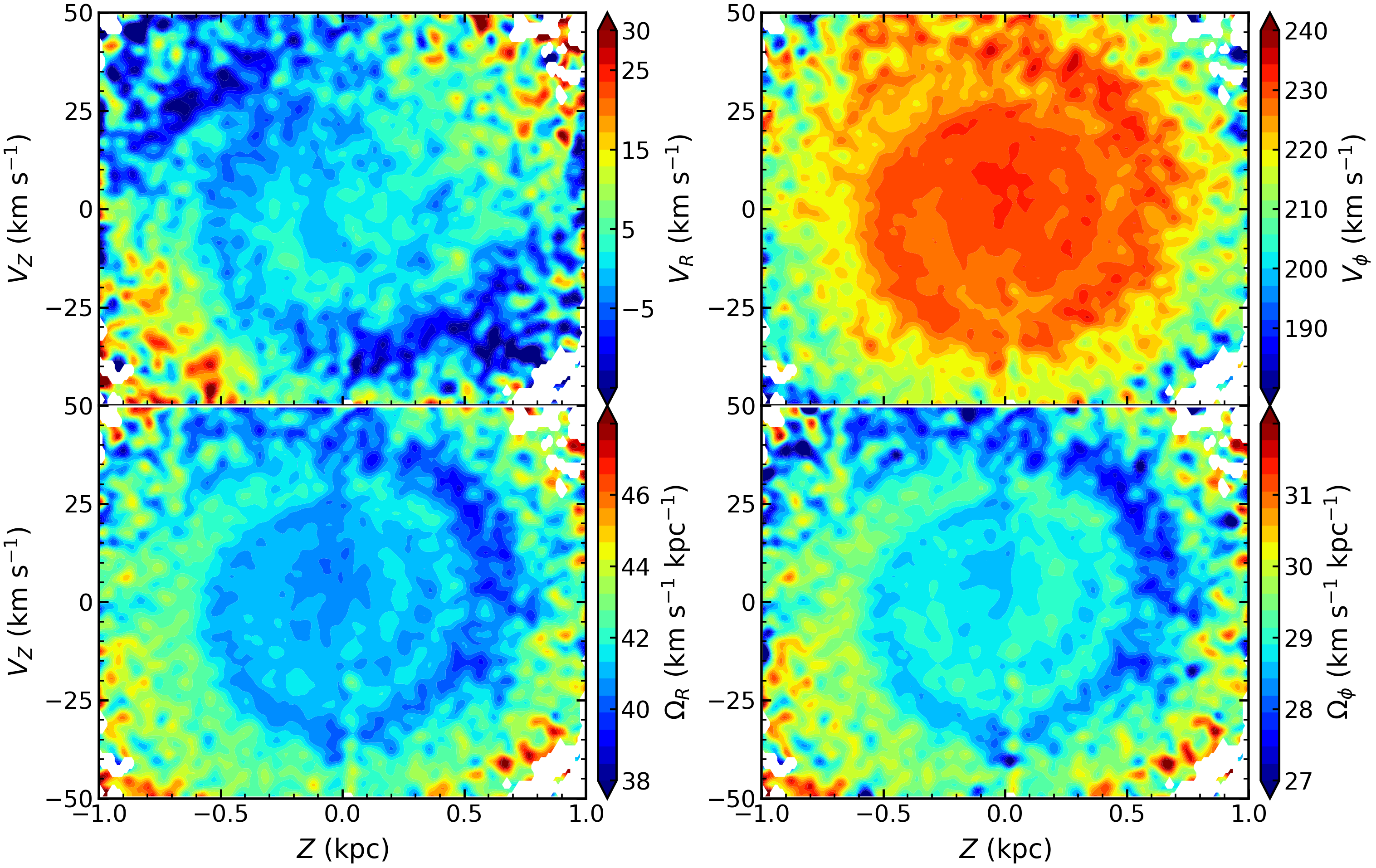}
\caption{\small Phase spiral of the GALAH DR4 disk sample in the solar-neighborhood, $7.2<R<9.2\,{\rm kpc}$. The upper-left, upper-right, lower-left, and lower-right panels are colour-coded by the median values of the radial velocity $v_R$, azimuthal velocity $v_\phi$, radial orbital frequency $\Omega_R$, and azimuthal orbital frequency $\Omega_\phi$, respectively. The phase-spiral morphology is schematically illustrated in Fig.~\ref{fig:GALAH_Same_Dynamical_Framework}.} 
\label{fig:GALAH_z_vz}
\end{figure*}

Following the analysis of the ridge structures, we examine vertical phase space to investigate how in-plane kinematics and orbital frequencies vary across the phase spiral. Fig.~\ref{fig:GALAH_z_vz} presents the GALAH DR4 disk sample in the vertical $z$--$v_z$ plane over the Galactocentric radial range $R=7.2$--$9.2\,\mathrm{kpc}$, with the maps colour-coded by $v_R$, $v_\phi$, $\Omega_R$, and $\Omega_\phi$.

These maps exhibit spiral-shaped variations, but their morphology and contrast differ. In the $v_R$ map, the spiral is primarily traced by regions of negative radial velocity and exhibits a two-armed morphology, similar to that shown by \citet{Antoja2023}. Our analysis complements this work by further examining maps of the same $(Z,v_z)$ plane coloured by the radial and azimuthal frequencies, $\Omega_R$ and $\Omega_\phi$. In the subsequent analysis, we extend our investigation to maps coloured by the orbital actions, $J_R$, $J_\phi$, and $J_z$, as well as to the phase spiral in different stellar age groups. These additional analyses allow us to compare the morphology and visibility of the phase spiral across a broader set of dynamical quantities and stellar age groups, providing complementary information on the orbital properties associated with the vertical phase-space structure. Two-armed phase spirals have also been identified by \citet{Alinder2024}, suggesting that the two-armed phase spiral is detectable only within a narrow distance range. The reason why the coloured azimuthal velocity is not clearly visible here will be addressed in a separate work; this may be related to a quadrupole phenomenon due to the velocity ellipsoid as mentioned in \citet{BlandHawthorn2019}.

The structure in the $v_\phi$ map is more tightly wound and is dominated by one-armed spiral. Similar one-armed structures are also visible in the $\Omega_R$ and $\Omega_\phi$ maps, with slightly stronger spiral contrast in $\Omega_R$ than in $\Omega_\phi$. The presence of spiral-like structures in both the velocity and frequency maps indicates that the phase-space spiral is associated with systematic variations in the in-plane kinematics and orbital frequencies of the sampled stars.

\begin{figure*}[!t]
\centering
\includegraphics[width=0.95\textwidth]{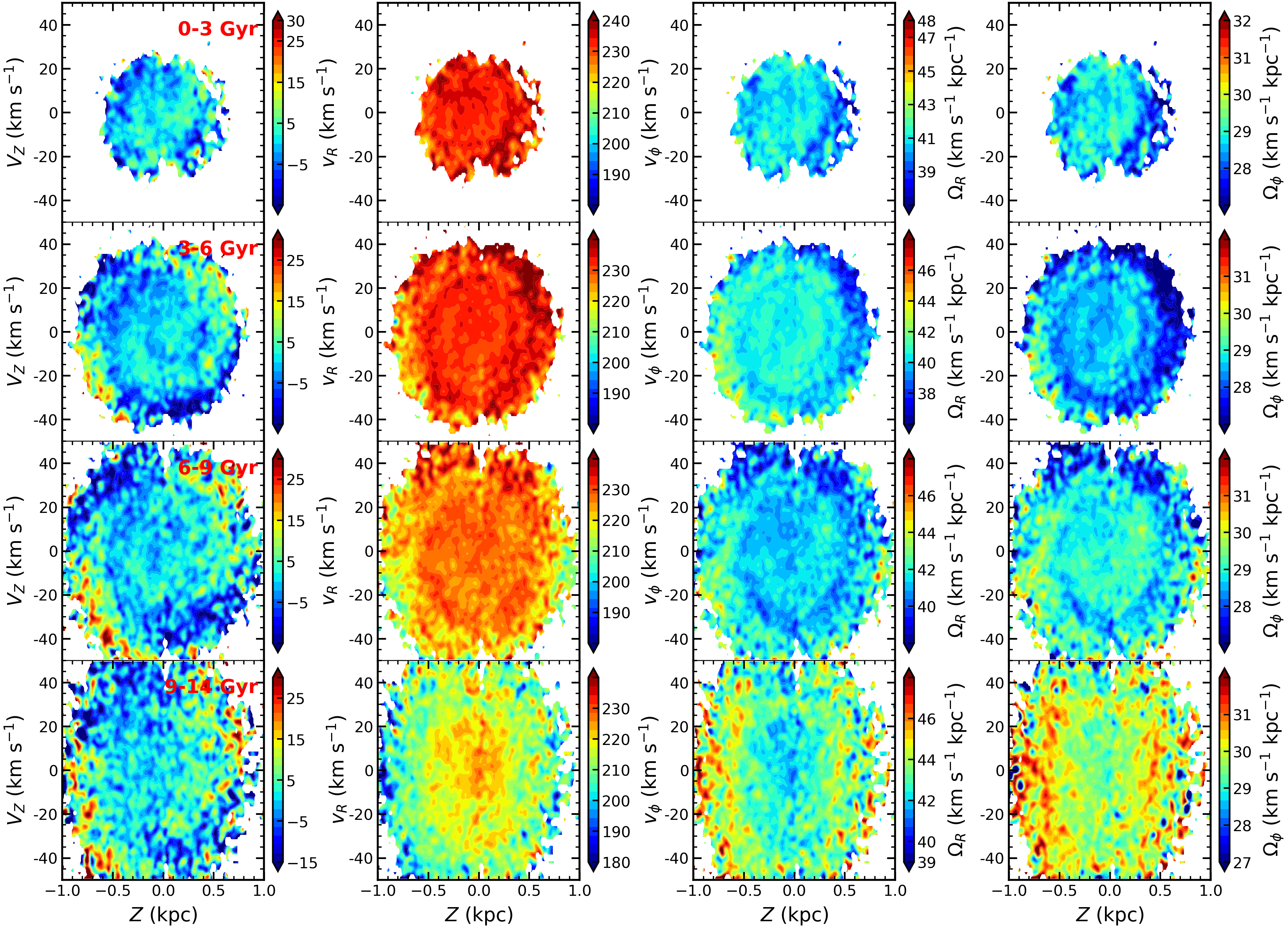}
\caption{\small Same as Fig.~\ref{fig:GALAH_z_vz}, but for different stellar-age bins. From top to bottom, the rows correspond to stars with ages of $0$--$3$, $3$--$6$, $6$--$9$, and $9$--$14\,{\rm Gyr}$. From left to right, the columns are color-coded by the median values of the radial velocity $v_R$, azimuthal velocity $v_\phi$, radial orbital frequency $\Omega_R$, and azimuthal orbital frequency $\Omega_\phi$, respectively. 
}
\label{fig:GALAH_z_vz_age}
\end{figure*}

To examine the age dependence of these spiral-like features, we divide the local sample into four stellar-age intervals and present the corresponding maps in Fig.~\ref{fig:GALAH_z_vz_age}. The spiral-like signatures are relatively weak in the youngest subsample, are clearest in the $3$--$6$ and $6$--$9$~Gyr bins, and have lower contrast in the oldest subsample. The age dependence of spiral winding is more apparent in the $\Omega_R$ and $\Omega_\phi$ maps than in the $v_R$ and $v_\phi$ maps: the phase spirals traced by younger stellar populations are more tightly wound, whereas those traced by older populations are more loosely wound.

\begin{figure*}[!t]
\centering
\includegraphics[width=0.95\textwidth]{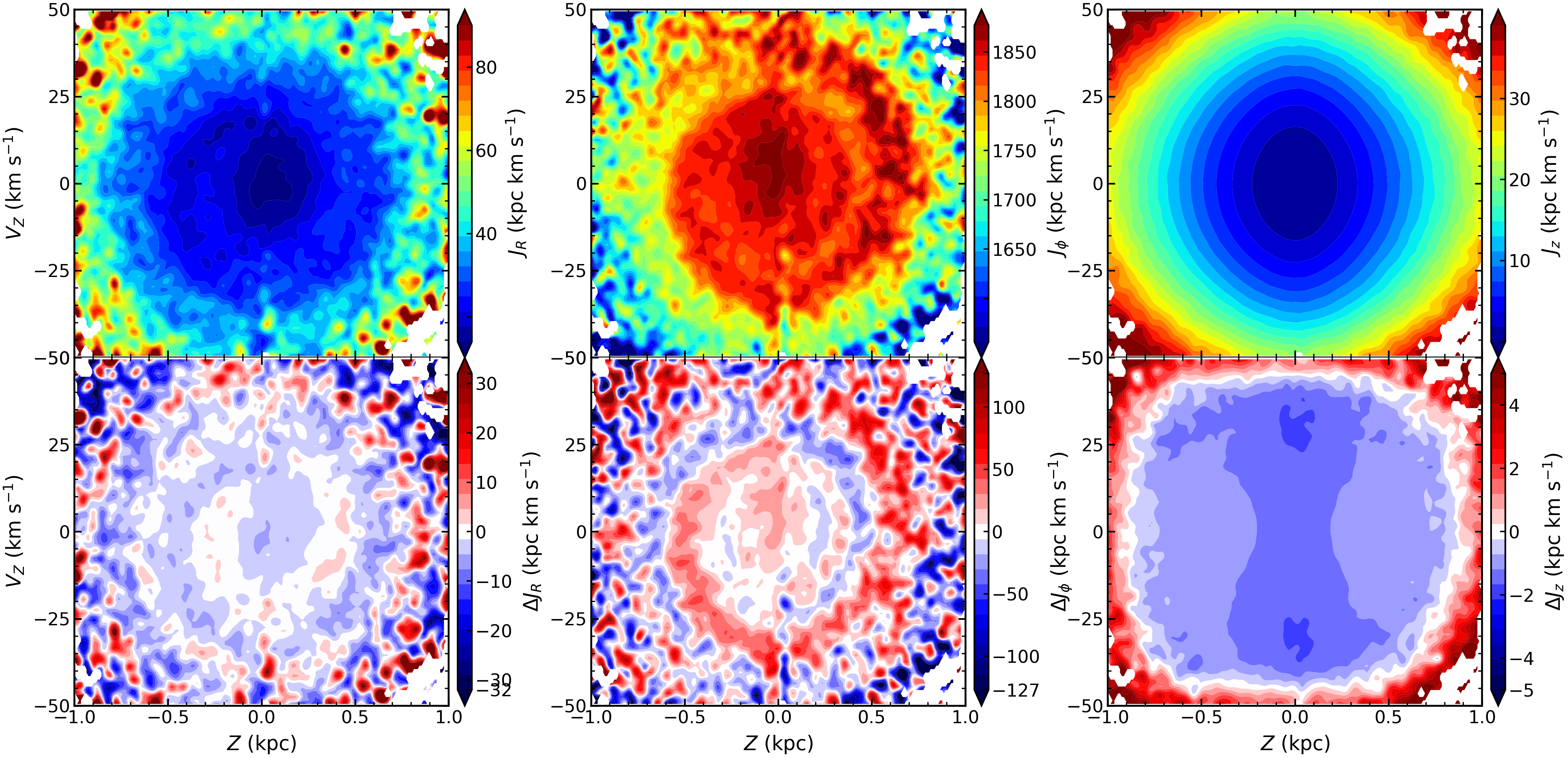}
\caption{\small Action signatures of the phase spiral in the GALAH DR4 sample within the solar-neighborhood, $7.2<R<9.2\,{\rm kpc}$. From left to right, the top-row panels are color-coded by the median radial, azimuthal, and vertical actions, $J_R$, $J_\phi$, and $J_z$, respectively. The bottom-row panels show the corresponding residual distributions, defined as $\Delta J_i=J_{i,\rm smooth}-J_{i,\rm bg}$, where $i\in{R,\phi,z}$ and $J_{i,\rm bg}$ represents the large-scale smoothed background of the corresponding action map. 
}
\label{fig:GALAH_Z_Vz_J_actions_residuals}
\end{figure*}

To examine whether the phase spiral is also reflected in the orbital-action distributions, we map the median values of $J_R$, $J_\phi\equiv L_z$, and $J_z$ in each pixel of the $z$--$v_z$ plane. Fig.~\ref{fig:GALAH_Z_Vz_J_actions_residuals} presents these action distributions and the corresponding residual maps obtained after subtracting the smooth, large-scale background.

Spiral-related variations are visible in both the $J_R$ and $J_\phi$ maps, although their coherence and contrast differ. The $J_\phi$ distribution exhibits a more clearly traced spiral-like pattern, while the $J_R$ map also contains a weaker but identifiable spiral signature. After subtracting the large-scale background, these non-axisymmetric features become more evident: the $\Delta J_\phi$ map reveals a prominent spiral structure, and a coherent spiral pattern is also visible in the $\Delta J_R$ map. By contrast, the $J_z$ distribution is dominated by a smooth increase from the centre of the $z$--$v_z$ plane toward larger vertical phase-space amplitudes, reflecting the overall dependence of vertical action on the amplitude of vertical motion or oscillation. After removal of the background, the $\Delta J_z$ map does not exhibit a comparable coherent spiral pattern during this work.

\begin{figure*}[!t]
\centering
\includegraphics[width=0.95\textwidth]{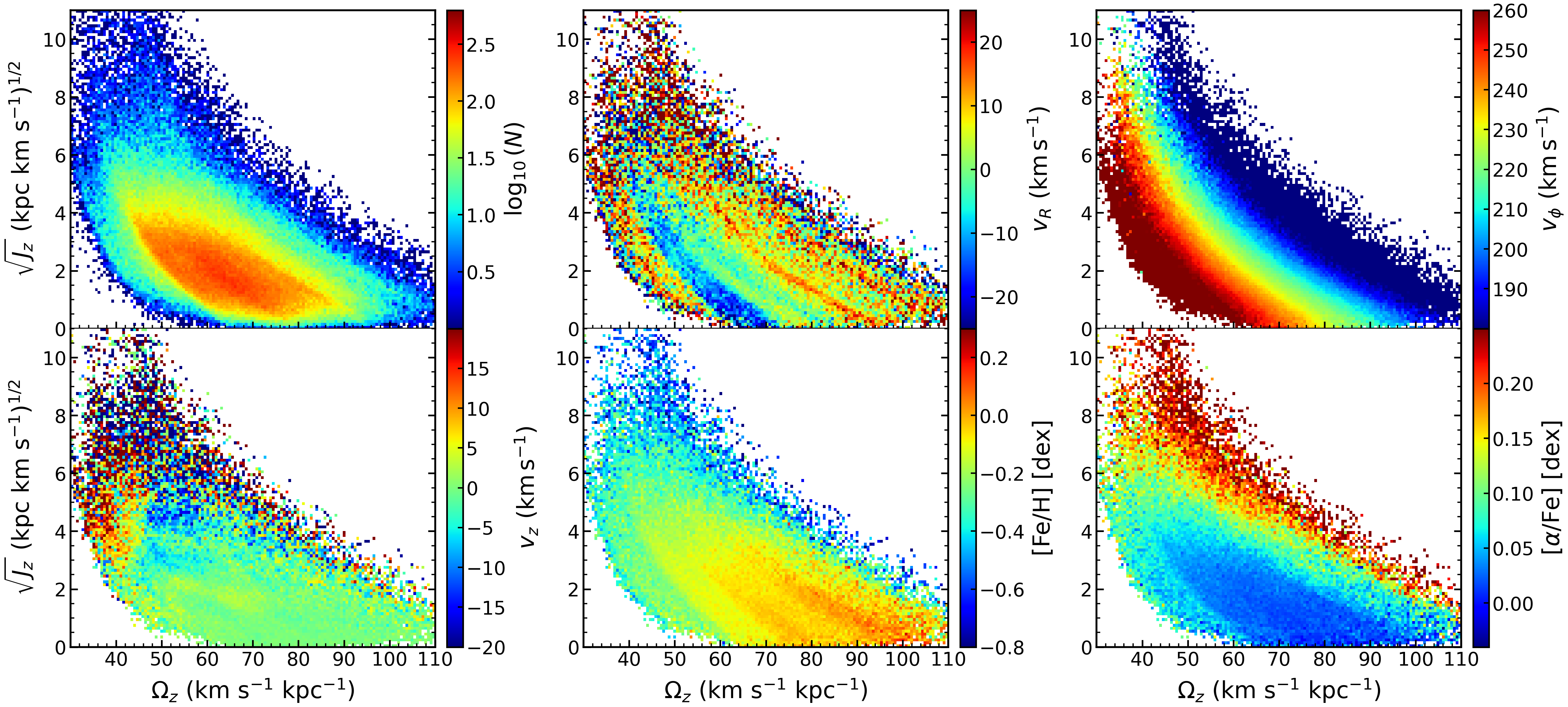}
\caption{\small Distribution of the local GALAH DR4 disk sample in the $\Omega_z$--$\sqrt{J_z}$ plane. The sample is restricted to the same solar-neighborhood adopted for the $z$--$v_z$ phase-spiral analysis. From left to right, the top-row panels show the logarithm of the number of stars per bin, $\log_{10}N$, the mean radial velocity $v_R$, and the mean azimuthal velocity $v_\phi$, respectively. The bottom-row panels are color-coded by the mean vertical velocity $v_z$, metallicity $[\mathrm{Fe/H}]$, and $\alpha$-element abundance $[\alpha/\mathrm{Fe}]$, respectively.}
\label{fig:GALAH_Omegaz_Jz}
\end{figure*}

To examine the frequency--action relation underlying vertical phase mixing, we investigate the distribution of the same local GALAH DR4 disk sample used in the phase-spiral analysis in the $\Omega_z$--$\sqrt{J_z}$ plane. Fig.~\ref{fig:GALAH_Omegaz_Jz} presents the corresponding maps for stars satisfying $7.2<R<9.2\,\mathrm{kpc}$, $|z|<1\,\mathrm{kpc}$, and $|v_z|<50\,\mathrm{km\,s^{-1}}$.

The logarithmic stellar-count map reveals a broad, curved band extending from high $\Omega_z$ and low $\sqrt{J_z}$ toward lower $\Omega_z$ and higher $\sqrt{J_z}$. This overall trend is consistent with the decrease in vertical oscillation frequency with increasing vertical action in an anharmonic disk potential \citep{Binney2018}.

Within this band, the $v_\phi$ map displays a smooth, systematic velocity gradient. For stars observed at comparable Galactocentric radii, $v_\phi$ is closely related to angular momentum through $L_z=Rv_\phi$. The observed gradient is therefore consistent with a dependence of $\Omega_z$ on angular momentum in addition to vertical action, as discussed by \citet{Binney2018}. The $v_R$ and $v_z$ maps also show localised variations between positive and negative velocities.

The chemical-abundance maps exhibit systematic large-scale gradients across the frequency--action distribution. Stars at low $\sqrt{J_z}$ are generally more metal-rich and low-$\alpha$, whereas those at higher $\sqrt{J_z}$ tend to be more metal-poor and $\alpha$-enhanced. These gradients indicate that stellar populations with different chemical properties also differ in their vertical-action distributions, linking chemical composition to vertical orbital properties in the selected sample.

\begin{figure*}[!t]
\centering
\includegraphics[width=0.98\textwidth]{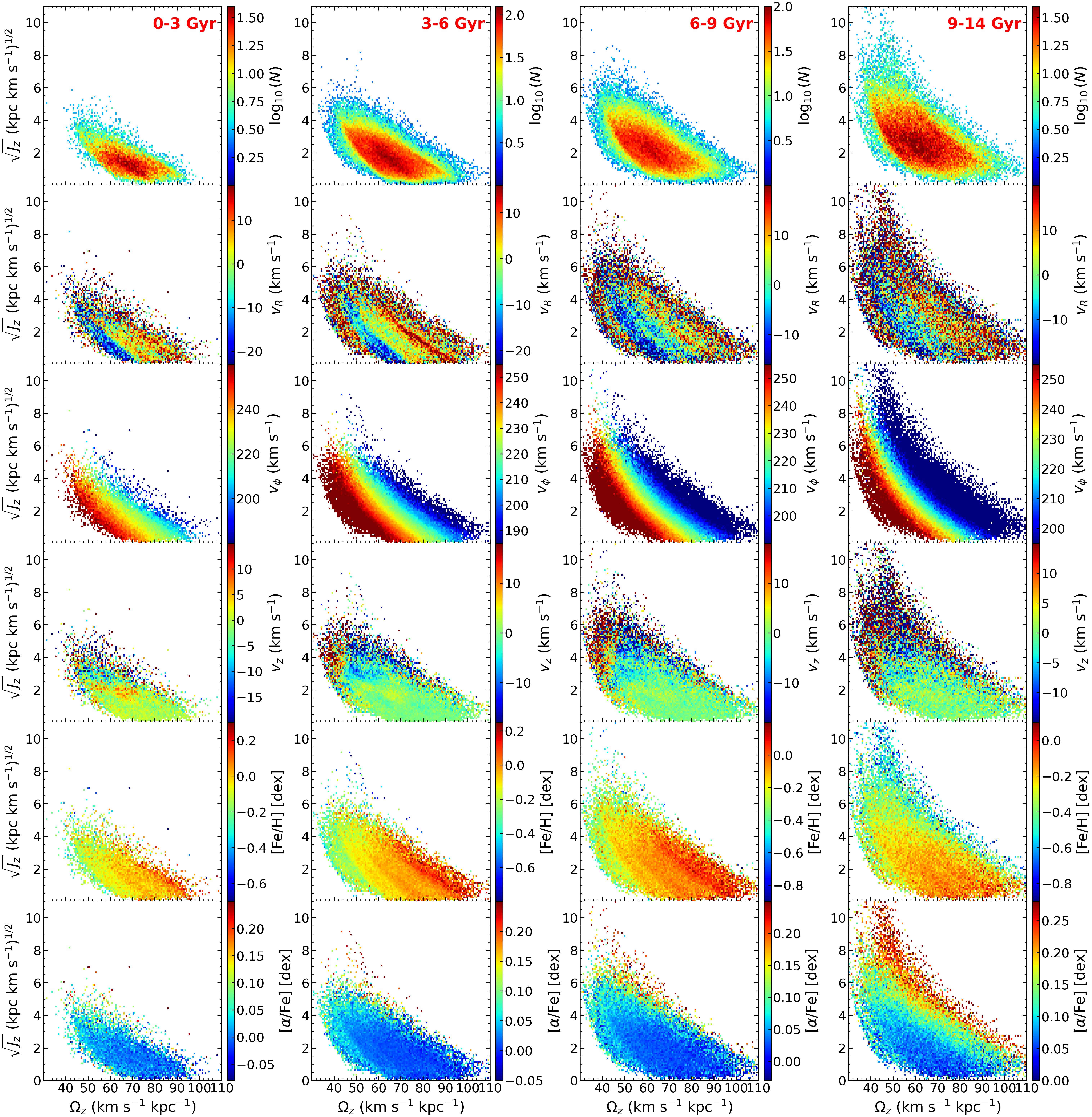}
\caption{\small Age dependence of the frequency--action structures for the local GALAH sample. The sample is restricted to $7.2<R<9.2\,{\rm kpc}$, $|z|<1\,{\rm kpc}$, and $|v_z|<50\,{\rm km\,s^{-1}}$. From left to right, the columns correspond to stellar ages of $0$--$3$, $3$--$6$, $6$--$9$, and $9$--$14\,{\rm Gyr}$. The top row shows the logarithm of the number of stars per bin, $\log_{10}N$. From top to bottom, the remaining rows show maps color-coded by $v_R$, $v_\phi$, $v_z$, $[\mathrm{Fe}/\mathrm{H}]$, and $[\alpha/\mathrm{Fe}]$, respectively.
}
\label{fig:GALAH_Omegaz_Jz_age}
\end{figure*}

In a similar vein, to examine the age dependence of the vertical frequency--action distribution, we compare the maps of the same local sample divided into four stellar-age bins in Fig.~\ref{fig:GALAH_Omegaz_Jz_age}. The youngest population is concentrated within a relatively narrow range of $\sqrt{J_z}$, whereas older populations span a broader range and extend to larger vertical actions. In the kinematic maps, $v_\phi$ is dominated by a smooth, systematic large-scale velocity gradient. This gradient is therefore consistent with stars of similar vertical action but different angular momenta having different vertical oscillation frequencies. By comparison, the $v_R$ and $v_z$ maps exhibit localised variations between positive and negative velocities, with stripe-like features in some regions. The stripe-like features in the $v_R$ maps are most clearly visible in the subsamples spanning $0$--$9$~Gyr and are less readily identifiable in the oldest subsample.

These differences indicate that stellar populations of different ages differ in both their vertical-action distributions and the visibility of their local kinematic features, which indicates that the heating and phase-mixing processes evolve over time.

The phase spiral is accompanied by systematic variations in the in-plane velocities and orbital frequencies, together with action signatures that are clearer in $J_\phi$ than in $J_R$. Its contrast and winding vary with stellar age, with the spiral most clearly visible in the $3$--$9$~Gyr populations. The $\Omega_z$--$\sqrt{J_z}$ maps reveal a smooth azimuthal-velocity gradient, localised variations in $v_R$ and $v_z$, and large-scale chemical-abundance gradients. Older populations occupy broader vertical-action ranges, highlighting the need to consider differences in the underlying orbital distributions when comparing the visibility of these features across stellar-age populations.

\section{Discussion}\label{sec:discussion}

\subsection{Coexistence of Ridges and Phase Spirals in a Common Phase-Space Framework}
\label{subsec:ridge_spiral_coexistence}

\begin{figure*}[htp!]
\centering
\includegraphics[width=0.95\textwidth]{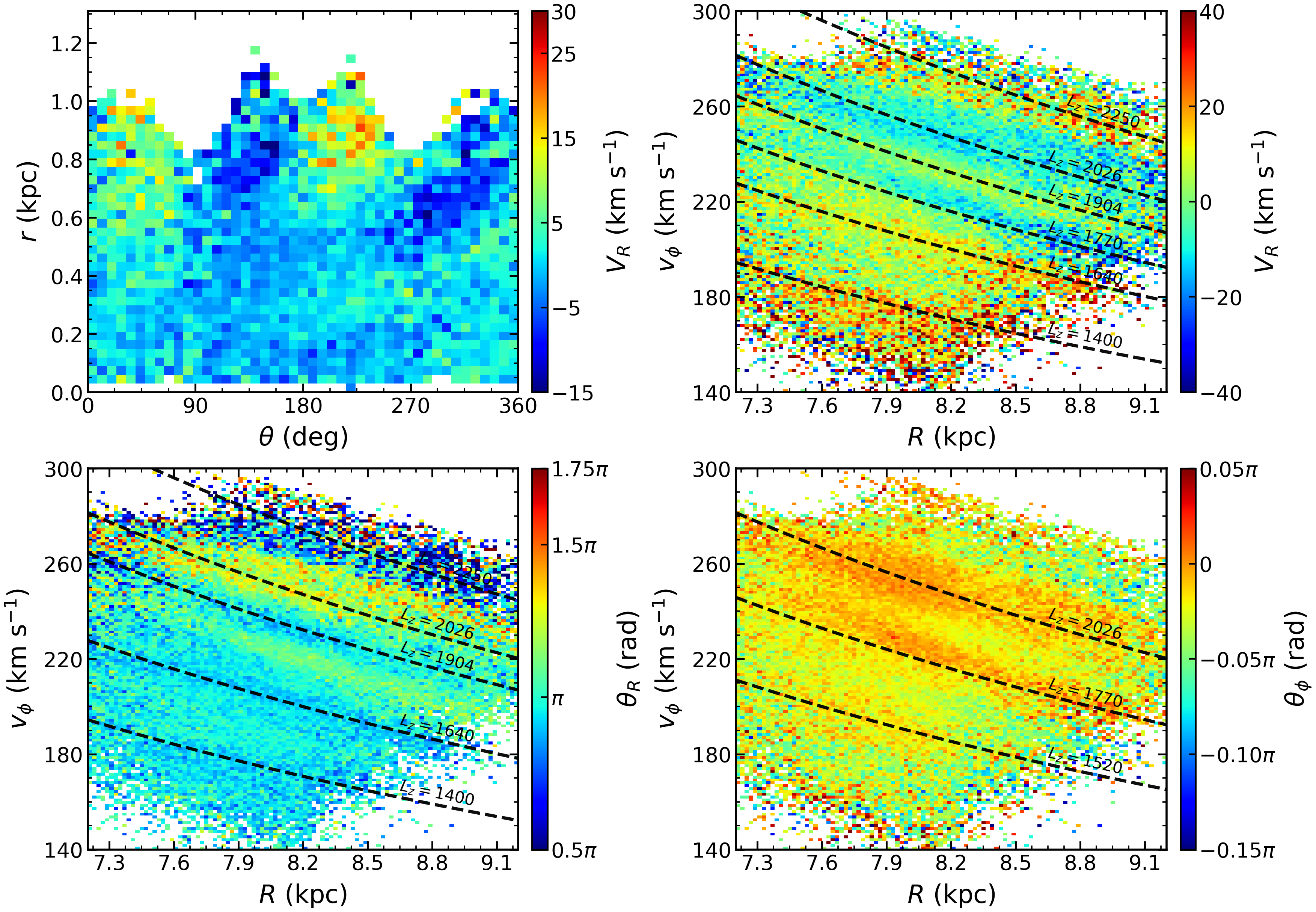}
\caption{\small Vertical phase-space structure and ridges in the GALAH DR4 sample, selected with the same spatial and kinematic cuts adopted for the phase-spiral analysis: \(7.2<R<9.2\,\mathrm{kpc}\), \(|z|<1\,\mathrm{kpc}\), and \(|V_z|<50\,\mathrm{km\,s^{-1}}\). The upper-left panel shows the mean radial velocity \(V_R\) in the \((\theta,r)\) plane, where \(r=[z^2+(V_z/S)^2]^{1/2}\), with \(S=60\,\mathrm{km\,s^{-1}\,kpc^{-1}}\), and \(\theta\) is measured counterclockwise from the positive \(z\)-axis in the \((z,V_z/S)\) plane. The upper-right, lower-left, and lower-right panels show distributions in the \(R\)–\(v_\phi\) plane, colour-coded by the mean radial velocity \(V_R\), mean radial angle \(\theta_R\), and mean azimuthal angle \(\theta_\phi\), respectively. }
\label{fig:GALAH_R_vphi_ridge_snail}
\end{figure*}

To further examine whether the in-plane ridges and the vertical phase spiral coexist within the same local stellar sample, we apply the selection criteria adopted for the phase-spiral analysis, namely $7.2<R<9.2\,{\rm kpc}$, $|z|<1\,{\rm kpc}$, and $|v_z|<50\,{\rm km\,s^{-1}}$, and present complementary views of the selected stars in Fig.~\ref{fig:GALAH_R_vphi_ridge_snail}. The upper-left panel shows the mean radial velocity in the unwrapped $(\theta,r)$ representation of vertical phase space, revealing inclined band-like structures associated with the phase spiral. The upper-right, lower-left, and lower-right panels show the $R$--$v_\phi$ distribution color-coded by the mean radial velocity $v_R$, radial orbital angle $\theta_R$, and azimuthal orbital angle $\theta_\phi$, respectively. Ridge-like features remain visible in these maps and broadly follow the overplotted constant-angular-momentum curves. Within the overlapping radial range, their locations, slopes, and overall morphology are broadly consistent with those of the corresponding structures in the larger-volume disk sample shown in Fig.~\ref{fig:GALAH_R_vphi_age}. Thus, the local stellar sample exhibiting the vertical phase spiral also retains identifiable in-plane ridges, which is consistent with the idea that they share the same underlying dynamical distribution function, but exhibit different responses in different phase-space or dynamical-space representations.

\begin{figure}[t]
    \centering
    \includegraphics[width=\columnwidth]{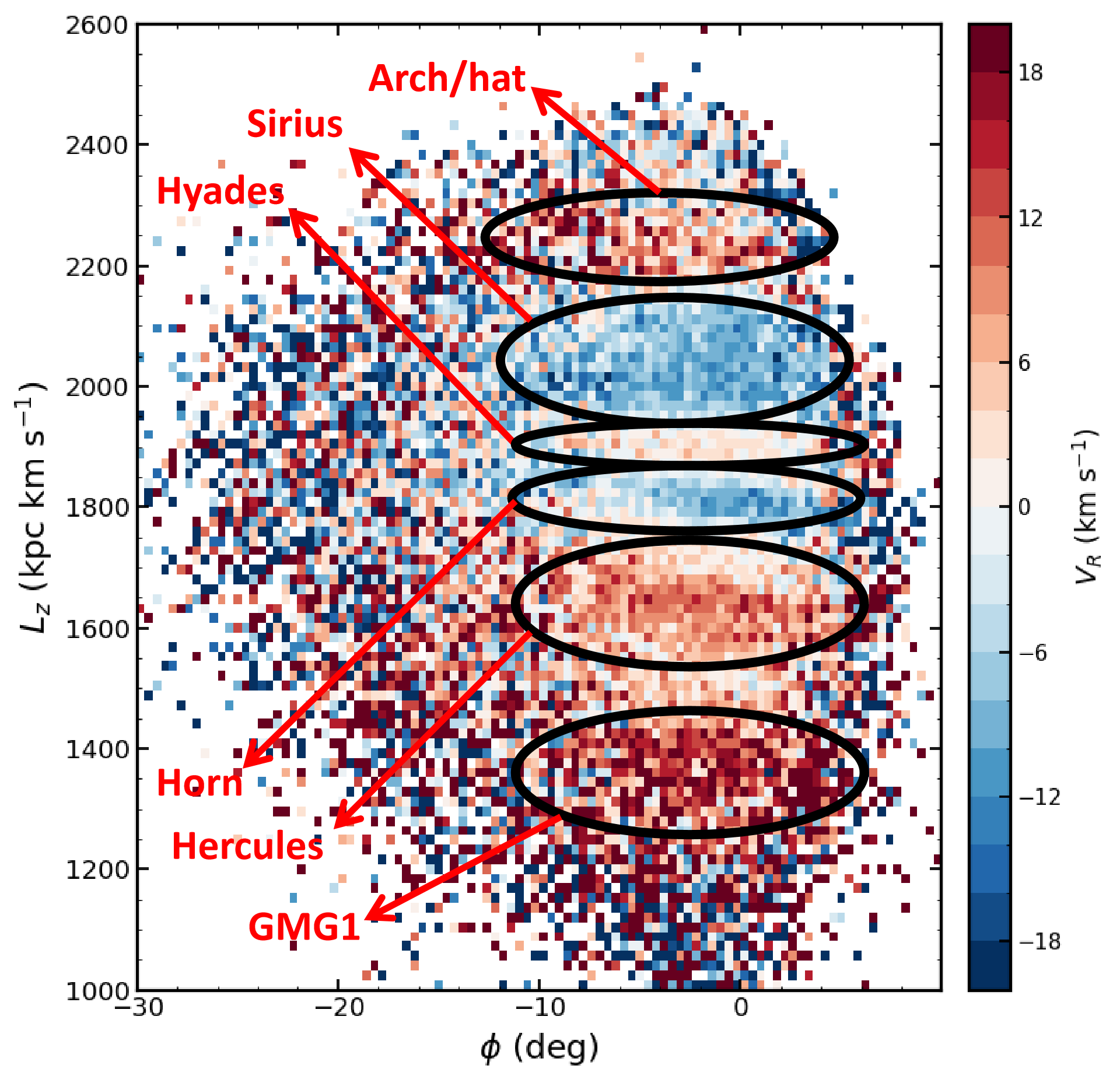}
    \caption{\small Mean Galactocentric radial velocity, $V_R$, in the $(\phi, L_z)$ plane for stars from GALAH DR4, selected within
$7.2 < R < 9.2\,\mathrm{kpc}$, $|z| < 1\,\mathrm{kpc}$, and
$|V_z| < 50\,\mathrm{km\,s^{-1}}$. Black circles highlight selected kinematic features, while red arrows and labels identify Arch/Hat, Sirius, Hyades, Horn, Hercules, and GMG1. Their respective characteristic $L_z$ values, in units of $\mathrm{kpc\,km\,s^{-1}}$, are $> 2200$, $\approx 2050$, $\approx 1880$, $\approx 1810$, $\approx 1615$, and $\approx 1385$. The moving-group regions are illustrated schematically in Fig.~\ref{fig:GALAH_Same_Dynamical_Framework}.}
\label{fig:GALAH_Moving_Group}
\end{figure}

Ridge structures are generally associated with features in the distributions of angular momentum and radial action, resonant orbital families, and radial migration, whereas phase spirals are commonly interpreted as signatures of vertical phase mixing following a perturbation \citep[e.g.][]{Antoja2018,Binney2018,Trick2019,Hunt2019,Antoja2023}. The simultaneous detection of both structures under identical spatial and kinematic selection criteria indicates that their coexistence cannot be simply attributed to differences in sample definition. It also shows that the local disk retains non-equilibrium structure in both its in-plane and vertical degrees of freedom, sharing a common dynamical distribution function but exhibiting different responses across different phase-space projections. An example of how structures in different projections are related within the in-plane dynamics was also provided by \citet{Hunt2020}, who showed that the ridges in guiding-radius and guiding-centre space in their Gaia sample correspond to some known moving groups in velocity space. As a test, when we use the same stellar sample to investigate whether moving-group signals can be detected, as shown in Figure~\ref{fig:GALAH_Moving_Group}, we clearly identify many ridge-like stars---in particular, the Hercules moving group (black circle)---which is consistent with the work of \citet{Monari2019}, who explained this feature in terms of bar resonances.In addition to Hercules, we identify the Arch/hat, Sirius, Hyades, Horn, and GMG1 velocity substructures, consistent with those reported by \citet{Li2025} and \citet{Lehmann2025}. \citet{Khoperskov2020} also noted that, by relocating stars to their guiding radius, the Milky Way spiral structure becomes visible as ridge-like overdensities in the \textit{Gaia} data, consistent with our findings here. This suggests that moving groups, ridges, and phase spirals may be connected within a common dynamical framework or common physical origins.

As discussed above, the detected ridges approximately follow curves of constant angular momentum. Comparing their locations with those of the moving groups marked in Fig.~\ref{fig:GALAH_Moving_Group}, we find that most of the ridges identified in Fig.~\ref{fig:GALAH_R_vphi_age} can be associated with one of these moving groups. Specifically, the ridges at $L_z = 2250$, $2026$, $1904$, $1770$, $1640$, and $1400\,\mathrm{kpc\,km\,s^{-1}}$ can be associated with Arch/Hat, Sirius, Hyades, Horn, Hercules, and GMG1, respectively.

\begin{figure*}[htp!]
\centering
\includegraphics[width=0.8\textwidth]{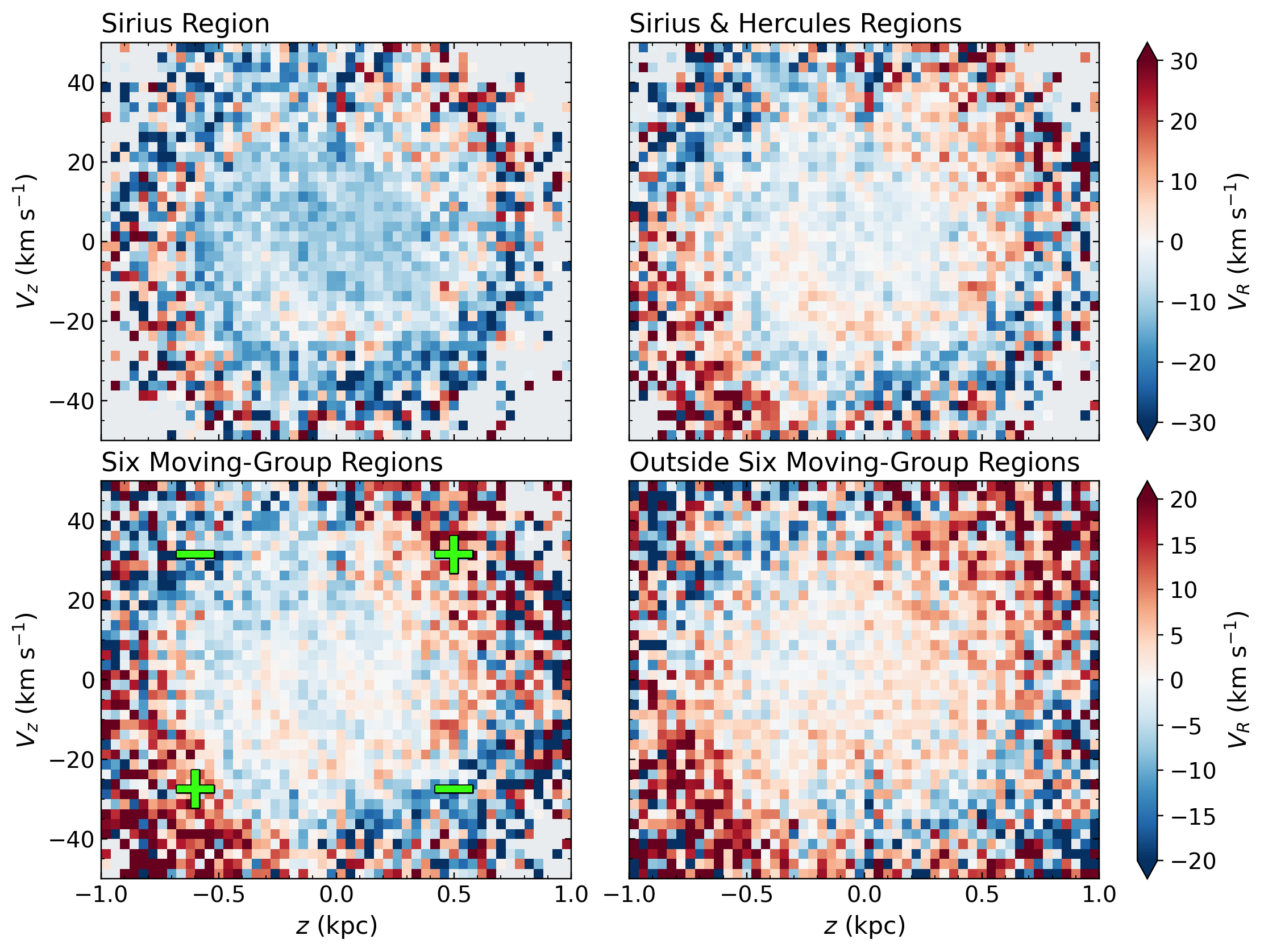}
\caption{\small Mean Galactocentric radial velocity, $V_R$, in the vertical phase plane $(z,V_z)$ for stars inside and outside the moving-group regions. All four panels are drawn from the same GALAH DR4 disc sample as Fig.~\ref{fig:GALAH_Moving_Group}. Six selections correspond to the kinematic structures highlighted in that figure, and a seventh comprises the stars that lie outside all six regions. The upper-left panel shows stars in the Sirius region, while the upper-right panel shows stars in the union of the Sirius and Hercules regions. The lower-left panel shows stars in the union of all six moving-group regions, while the lower-right panel shows the complementary sample. The white $+$ and $-$ symbols in the lower-left panel mark the positive and negative regions of the quadrupolar pattern in $V_R$. This pattern may be associated with the tilt of the velocity ellipsoid, as discussed by \citet{BlandHawthorn2019}.
}
\label{fig:GALAH_Moving_Group_phase_spiral_quadrupole}
\end{figure*}

Having identified a correspondence in angular momentum between ridges and moving groups in the Galactic disk, we next examine whether these groups affect the observed properties of the vertical phase spiral. Using the GALAH DR4 disk sample shown in Fig.~\ref{fig:GALAH_Moving_Group} as our parent sample, we select stars in the $(\phi,L_z)$ regions associated with the different moving groups and compare their maps of the mean radial velocity $V_R$ in the vertical phase plane $(z,V_z)$. As shown in Fig.~\ref{fig:GALAH_Moving_Group_phase_spiral_quadrupole}, a clear phase spiral is visible in the Sirius region alone, in the union of the Sirius and Hercules regions, and in the union of six moving-group regions. This seems to suggest that moving groups may make the phase-spiral structure more prominent. After excluding the union of the six regions from the parent sample, the spiral-like pattern is difficult to discern among the remaining stars. This comparison suggests a link between moving-group substructure and the observed phase spiral.

\begin{figure*}[htp!]
\centering
\includegraphics[width=0.7\textwidth]{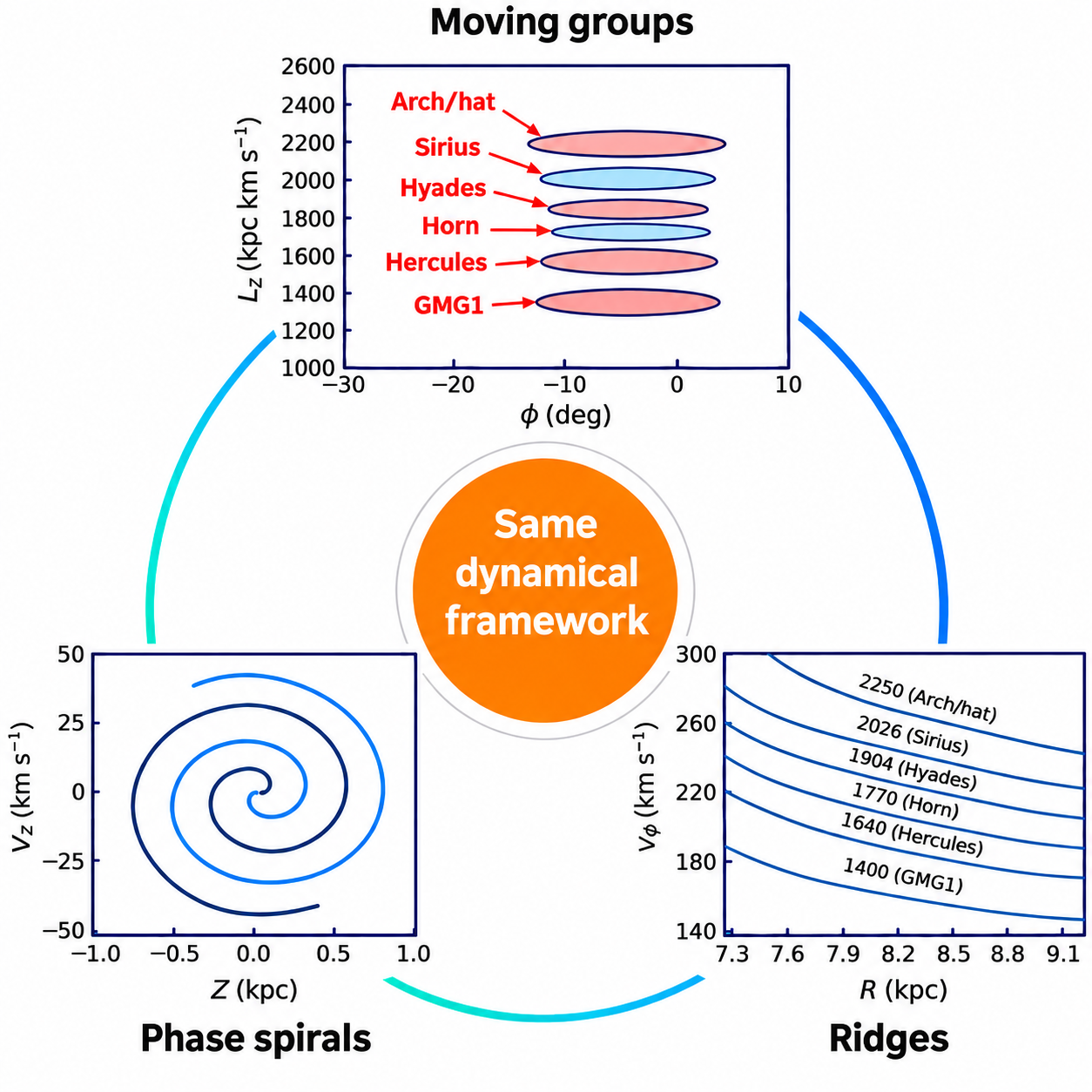}
\caption{\small Schematic illustration of the connection between moving groups, ridges, and phase spirals in the Galactic disc. The upper panel illustrates the distribution of six moving groups in the $\phi$--$L_z$ plane (Fig.~\ref{fig:GALAH_Moving_Group}), while the lower-right and lower-left panels represent the ridge structures in the $R$--$v_\phi$ plane (Fig.~\ref{fig:GALAH_R_vphi_age}) and the phase spiral in the $z$--$v_z$ plane (Fig.~\ref{fig:GALAH_z_vz}), respectively. These structures share a common dynamical framework.} 
\label{fig:GALAH_Same_Dynamical_Framework}
\end{figure*}

The correspondence in angular momentum between ridges and moving groups, together with the dependence of the phase spiral's visibility on the moving-group selection, indicates an observational link among these three kinematic substructures. Figure~\ref{fig:GALAH_Same_Dynamical_Framework} summarizes these connections schematically, showing the moving groups in the $(\phi,L_z)$ plane, the phase spiral in the $(z,V_z)$ plane, and the ridges in the $(R,V_\phi)$ plane. We therefore propose that these structures can be understood within a common dynamical framework as signatures of Galactic disk dynamics in different phase-space projections.

\subsection{Action-space clues to the clear connection between ridges and phase spirals}
\label{subsec:ridge_spiral_action_connection}

In the preceding section, we have identified the connection between ridges and phase spirals, interpreting them as manifestations of the same stellar phase-space distribution in different projections. This motivates a closer examination of their dynamical relationship, to this end, we investigate their dependence on orbital actions, angular momentum, and orbital phase, as discussed below.

The ridges identified in this study are closely associated with the in-plane orbital distribution of disk stars. In the $R$--$v_\phi$ plane, they are  aligned with curves of constant angular momentum and exhibit coherent patterns in the radial and azimuthal orbital angles, $\theta_R$ and $\theta_\phi$. In the $L_z$--$J_R$ plane, the ridges form diagonal sequences, while related structures are also identifiable in the $v_R$, $\theta_R$, and $\theta_\phi$ distributions in the $L_z$--$E$ plane. These results indicate that the ridges are jointly organised by angular momentum, radial action, and non-uniform distributions of the radial and azimuthal orbital phases.

Meanwhile, when the $z$--$v_z$ plane is colour-coded by $J_R$ and $J_\phi$, variations associated with the phase spiral are visible in both action maps and remain identifiable after subtraction of the smooth, large-scale background. The spiral signal is particularly clear in $J_\phi$, whereas the corresponding variations in $J_R$ are relatively weak. Since $J_R$ traces the amplitude of radial orbital oscillations and $J_\phi$ is the azimuthal action, these patterns indicate that stars at different locations along the phase spiral have different in-plane orbital properties.

These results link ridges and phase spirals through a common set of in-plane orbital variables. The ridges follow constant-angular-momentum curves and are closely associated with radial action and non-uniform orbital-phase distributions, while the phase spiral exhibits systematic variations in the same in-plane actions, $J_R$ and $J_\phi\equiv L_z$. This connection can be further understood through the guiding radius, which satisfies $L_z=R_gV_c(R_g)$ in an axisymmetric reference potential. Selecting stars within guiding-radius intervals therefore also selects corresponding angular-momentum intervals. In this context, \citet{Xu2026} identified clear two-armed $v_R$ phase spirals in several guiding-radius intervals and showed that their morphology depends on radial orbital phase, eccentricity, and contributions from different moving groups without detailed actions-age analysis. 

\citet{Antoja2018} revealed prominent ridges in the distribution of Galactocentric radius and azimuthal velocity, while \citet{Trick2019} characterized the rich substructure in the distribution of stellar orbital actions. Subsequently, \citet{Gaia2023Chem} identified action-space ridges associated with variations in metallicity and maximum orbital height, and \citet{Khanna2019} showed that ridge populations preferentially have higher metallicities than surrounding stars and approximately solar $\alpha$-element abundance ratios. In comparison with these recent advances, this study employs the publicly available GALAH DR4 catalog, we recover these features and connect their approximately constant-angular-momentum tracks in the $R$--$v_\phi$ plane with their signatures in the $z_{\max}$-coloured $J_\phi$--$J_R$ plane. We further identify coherent variations in the radial and azimuthal orbital angles, $\theta_R$ and $\theta_\phi$, associated with the same ridges, adding information on where their constituent stars lie along their orbits. This comparison complements the action-space characterization by \citet{Trick2019} and the use of orbital-angle structure to diagnose bar resonances by \citet{Trick2022}, to some extent. By tracing corresponding features across these distributions within a common sample, our analysis attempts to connect the ridges' angular momentum, radial excursions, vertical extent, and orbital phases in the space–time distribution. This provides additional observational constraints on their dynamical structure and its relationship to stellar populations and temporal evolution, thereby establishing a foundation for detailed comparisons with future data and simulations.

Using GALAH and Gaia, \citet{BlandHawthorn2019} recovered the spiral in maps of both azimuthal and radial velocity and examined its dependence on stellar age, chemistry, actions, and location. \citet{Hunt2022} subsequently revealed changes in spiral morphology by grouping stars according to their angular momenta and azimuthal orbital phases, while \citet{Frankel2023} quantified its amplitude and winding in vertical frequency--angle space. Using Gaia DR3 data, \citet{Antoja2023} presented maps of the $z$--$v_z$ plane colour-coded by median radial velocity, showing a two-armed phase-spiral morphology. Drawing on simulations of satellite perturbations, they suggested that the spiral structure in the radial-velocity maps could reflect the combined effects of in-plane density waves and vertical waves excited by the same perturbation. Building on these studies, we compare the same GALAH DR4 $z$--$v_z$ distribution colour-coded by different orbital quantities in our work: the azimuthal velocity $v_\phi$ and the radial and azimuthal orbital frequencies, $\Omega_R$ and $\Omega_\phi$, reveal one-armed spirals; the radial velocity $v_R$ displays a two-armed morphology; and residual spiral signatures remain visible in the radial-action and angular-momentum maps, $J_R$ and $J_\phi$. These comparisons trace the spiral's dependence on in-plane orbital properties through corresponding features in velocity, frequency, and action maps with clear time evolution. In general, this behaviour is compatible with the interpretation of \citet{Binney2018}, who showed that a perturbation can affect both vertical and in-plane motions and emphasized that the vertical oscillation frequency depends on angular momentum as well as vertical oscillation amplitude, helping explain the spiral's visibility in $v_\phi$. More generally, \citet{Tremaine2023} demonstrated that a superposition of stochastic, spatially coherent perturbations can generate a phase spiral, while scattering by small-scale structure limits its survival. Simulations including a multiphase interstellar medium by \citet{TepperGarcia2025} also produced spirals visible in both $v_R$ and $v_\phi$, including in an isolated model, and explicitly described the linked vertical and in-plane signatures as ``kinematic interlocking''. It appears that our results, at this stage, are consistent with these modelling studies concerning kinematic interlocking, or the coupling between ridges and snails, although a more detailed comparison and a fuller assessment of the contributions from different mechanisms require further work. 

We now compare the morphology of the phase spiral in maps colour-coded by the radial and azimuthal orbital frequencies and the corresponding actions. This provides additional insight into its age dependence and its relationship with in-plane orbital properties.
Together with the evolution pattern in actions and orbital angles identified in our sample, these findings provide clues to a connection between the phase spiral and the in-plane orbital distribution traced by the ridges. From a dynamical perspective, a single perturbation can both alter stellar orbital energies and angular momenta, reshaping the in-plane orbital distribution, and drive the vertical phase distribution away from equilibrium. In the disk plane, phase mixing or resonant responses can produce ridges, some of which approximately follow constant-angular-momentum curves \citep{Khanna2019}. In the vertical direction, the dependence of the oscillation frequency $\Omega_z$ on $J_z$, $J_\phi$, and $J_R$ causes stars on different orbits to accumulate vertical phase at different rates, gradually winding the perturbed distribution into a phase spiral \citep{Binney2018,Antoja2023}. We therefore propose that ridges and phase spirals may share a common perturbation origin, tracing the in-plane and vertical evolution of the same perturbed stellar distribution. 

\subsection{Age and chemical dependence of the disk response}
\label{subsec:age_chemical_dependence}

The preceding analysis has explicitly revealed connections between ridges and phase spirals in terms of orbital actions, angular momentum, and orbital phase. Building on these results, examining their dependence on stellar age and chemical abundance helps clarify how these structures manifest in different stellar populations, time-scale picture, and provides further constraints on the population dependence of the disk dynamical response. Stellar populations of different ages occupy different regions of action space and exhibit different orbital properties and degrees of phase mixing, while chemical abundances also provide insight into distinct disk populations and their evolutionary and perturbation histories.

The ridge structures exhibit a dependence on stellar age. Across several age intervals, they occupy similar locations and follow the same constant-angular-momentum curves, although their contrast varies. Their detectability over a broad age range may be associated with long-lived dynamical structures or may reflect the repeated influence of certain dynamical processes on disk stellar orbits. The phase spiral likewise shows a pronounced age dependence. In the youngest stellar population, the spiral-related velocity and frequency structures are relatively weak and limited in extent. These structures are clearest in the $3$--$6\,{\rm Gyr}$ and $6$--$9\,{\rm Gyr}$ subsamples, whereas among the oldest stars, the spiral is more loosely wound and has substantially lower contrast.

The chemical-abundance maps reveal population differences that parallel the age dependence described above. In the $L_z$--$J_R$ plane, some ridge regions occupied by relatively metal-rich and low-$\alpha$ stellar populations \citep{Khanna2019,Khoperskov2022,Gaia2023Chem}. This is broadly consistent with previous chemical-mapping studies, which found that stars in many, but not all, ridge structures are more metal-rich than those in their local phase-space surroundings, and that some prominent ridges are associated with relatively low $[\alpha/{\rm Fe}]$. A corresponding chemical-abundance trend is also evident in the $\Omega_z$--$\sqrt{J_z}$ plane. Stars at lower $\sqrt{J_z}$ are generally more metal-rich and low-$\alpha$, whereas those at higher $\sqrt{J_z}$ are relatively metal-poor and $\alpha$-enhanced. Dynamically colder, metal-rich, and low-$\alpha$ populations exhibit clearer kinematic structures, while dynamically hotter populations show smoother distributions. We defer the investigation of this intriguing connection between star formation history and dynamics to a subsequent paper in our series.

Overall, these results show that the visibility of ridges and phase spirals is strongly population-age dependent. Their age and chemical-abundance trends may reflect differences in the orbital distributions and degrees of phase mixing of the underlying stellar populations. Combined with the action-space connections discussed above, these findings are consistent with a common dynamical origin. Their present-day morphology and contrast may also have been shaped by the combined long-term effects of multiple perturbations and subsequent orbital evolution. In future work, we will use quantitative dynamical simulations to assess this interpretation and the contributions of the different processes.

\section{Conclusions}
\label{sec:conclusion}

In this work, we used GALAH DR4 to investigate ridge structures and phase spirals within a common chemodynamical framework. We characterised their distributions in phase space, action and angle space, frequency--action space, and chemical-abundance maps, and examined their dependence on stellar age. By applying identical local spatial and kinematic selection criteria to both classes of substructure, we compared their properties within the same stellar population and discussed their origins. Our main conclusions are summarised as follows:

\begin{enumerate}

\item Multiple prominent ridges are detected in the $R$--$v_\phi$ plane, following curves of constant angular momentum.  Across several age intervals, the ridges occupy similar locations and follow the same constant-angular-momentum curves, although their contrast varies with stellar age and has time evolution picture.

\item Related structures are also identified in the $L_z$--$J_R$ and $L_z$--$E$ planes. In the $L_z$--$J_R$ plane, the ridge-like sequences are particularly clear in the $z_{\rm max}$ distribution. The ridge regions are characterised by metal-rich and $\alpha$-poor stellar populations. In the $L_z$--$E$ plane, corresponding structures appear in the velocity and orbital-angle distributions, most prominently in $v_R$, $\theta_R$, and $\theta_\phi$.

\item A prominent phase spiral is detected in the local $z$--$v_z$ plane. The $v_R$ distribution exhibits a two-armed morphology, whereas the $v_\phi$, $\Omega_R$, and $\Omega_\phi$ maps are dominated by one-armed spiral-like variations. When orbital actions are mapped onto the same plane, spiral-related signatures are visible in both $J_R$ and $J_\phi$ and remain identifiable after subtraction of the large-scale background. By contrast, the $J_z$ distribution is dominated by a smooth increase with vertical phase-space amplitude. The spiral features exhibit an age dependence and are most prominent in the $3$--$6\,{\rm Gyr}$ and $6$--$9\,{\rm Gyr}$ subsamples. In the $\Omega_R$ and $\Omega_\phi$ age maps, the spiral becomes more loosely wound with increasing stellar age.

\item In the $\Omega_z$--$\sqrt{J_z}$ plane, the solar-neighborhood sample forms a continuous curved band, with the vertical oscillation frequency generally decreasing as the vertical action increases. The smooth $v_\phi$ gradient across the band is consistent with $\Omega_z$ depending on both the vertical action $J_z$ and the angular momentum $L_z$. The band also exhibits systematic chemical-abundance gradients, from relatively metal-rich stars with lower $[\alpha/\mathrm{Fe}]$ at low $\sqrt{J_z}$ to relatively metal-poor stars with higher $[\alpha/\mathrm{Fe}]$ at high $\sqrt{J_z}$. The youngest population occupies a relatively narrow range of $J_z$, and older populations span a broader range extending to larger $J_z$. Variations in $v_R$ and $v_z$ are predominantly localized. The stripe-like features in $v_R$ are clearer in the age bins spanning $0$--$9\,\mathrm{Gyr}$ and are harder to identify in the oldest bin ($9$--$14\,\mathrm{Gyr}$).

\item In the same GALAH DR4 disk sample, the ridges follow curves of constant angular momentum and correspond in $L_z$ to the Arch/Hat, Sirius, Hyades, Horn, Hercules, and GMG1 moving groups. The vertical phase spiral is clearly visible in the Sirius region alone, in the combined Sirius and Hercules regions, and in the union of all six moving-group regions. After these six regions are excluded, the spiral is difficult to discern in the remaining sample. The correspondence in $L_z$ and the dependence of phase-spiral visibility on sample selection provide an observational basis for studying these three structures within a common dynamical framework.

\end{enumerate}

Taken together, our results support interpreting ridges, phase spirals, and moving groups as related manifestations of the same stellar phase-space distribution in different two-dimensional projections. The ridges trace coherent structure in angular momentum, radial action, and orbital phase, and the properties of the phase spiral vary systematically with the in-plane actions $J_R$ and $J_\phi \equiv L_z$. Some ridges correspond in angular momentum to moving groups, and the visibility of the phase spiral varies with the selection of moving-group regions. These observations connect the vertical phase-space structure to the in-plane orbital distribution and suggest that all three structures may be described by a common dynamical distribution function.

The dependence of ridges and phase spirals on stellar age and chemical abundance further indicates that their observed properties vary across stellar populations. Within this framework, the links identified above among ridges, phase spirals, and moving groups suggest that their formation and evolution may reflect the combined long-term effects of multiple perturbations. Non-axisymmetric evolution within the Galactic disk and external perturbations may both have contributed to shaping their present-day morphology. In future work, we will use dynamical simulations to investigate a possible common dynamical origin of the three structures and assess the roles of different perturbations in their formation and evolution.

\begin{acknowledgments}
This work was supported by Natural Science Foundation of Sichuan Province (Grant No. 2026YFTX0025) and the National Natural Science Foundation of China (NSFC) under grant 12373035. Y.P. was supported by Natural Science Foundation of Sichuan Province (Grant No. 2026YFTX0019). H.F.W. acknowledges financial support from the Overseas Talent Program. We appreciate the comments provided by Thor Tepper-Garcia and Martin Lopez-Corredoira.

This work made use of the Fourth Data Release of the GALAH Survey \citep{Buder2025}. The GALAH survey is based on observations made at the Australian Astronomical Observatory, under programmes A/2013B/13, A/2014A/25, A/2015A/19, A/2017A/18. We acknowledge the traditional owners of the land on which the AAT stands, the Gamilaraay people, and pay our respects to elders past and present. This paper includes data that have been provided by AAO Data Central (\url{https://datacentral.org.au/}).

This work has made use of data from the European Space Agency (ESA) mission \textit{Gaia} (\url{https://www.cosmos.esa.int/gaia}), processed by the \textit{Gaia} Data Processing and Analysis Consortium (DPAC, \url{https://www.cosmos.esa.int/web/gaia/dpac/consortium}). Funding for the DPAC has been provided by national institutions, in particular the institutions participating in the \textit{Gaia} Multilateral Agreement.

\end{acknowledgments}


\begin{thebibliography}{}

\bibitem[Antoja et al.(2018)]{Antoja2018}
Antoja, T., Helmi, A., Romero-G{\'o}mez, M., et al. 2018, Nature, 561, 360

\bibitem[Antoja et al.(2023)]{Antoja2023}
Antoja, T., Ramos, P., Garc{\'i}a-Conde, B., Bernet, M.,
Laporte, C.~F.~P., \& Katz, D. 2023,
A\&A, 673, A115

\bibitem[Alinder et al.(2024)]{Alinder2024}
Alinder, S., McMillan, P.~J., \& Bensby, T. 2024,
A\&A, 690, A15

\bibitem[Binney \& Tremaine(2008)]{Binney2008}
Binney, J. \& Tremaine, S. 2008,
Galactic Dynamics: Second Edition
(Princeton, NJ: Princeton Univ. Press)

\bibitem[Binney(2012)]{Binney2012}
Binney, J. 2012,
MNRAS, 426, 1324

\bibitem[Binney \& Sch{\"o}nrich(2018)]{Binney2018}
Binney, J. \& Sch{\"o}nrich, R. 2018,
MNRAS, 481, 1501

\bibitem[Bland-Hawthorn \& Gerhard(2016)]{BlandHawthorn2016}
Bland-Hawthorn, J. \& Gerhard, O. 2016,
ARA\&A, 54, 529

\bibitem[Bland-Hawthorn et al.(2019)]{BlandHawthorn2019}
Bland-Hawthorn, J., Sharma, S., Tepper-Garc{\'i}a, T.,
Binney, J., Freeman, K.~C., Hayden, M.~R., et al. 2019,
MNRAS, 486, 1167

\bibitem[Bland-Hawthorn \& Tepper-Garc{\'i}a(2021)]{BlandHawthorn2021}
Bland-Hawthorn, J. \& Tepper-Garc{\'i}a, T. 2021,
MNRAS, 504, 3168

\bibitem[Bovy(2015)]{Bovy2015}
Bovy, J. 2015,
ApJS, 216, 29

\bibitem[Buder et al.(2025)]{Buder2025}
Buder, S., Kos, J., Wang, X.~E., McKenzie, M., Howell, M.,
Martell, S.~L., Hayden, M.~R., Zucker, D.~B., Nordlander, T.,
Montet, B.~T., et al. 2025,
PASA, 42, e051

\bibitem[Cao et al.(2024)]{Li2024}
Cao, C., Li, Z.-Y., Sch{\"o}nrich, R.,
\& Antoja, T. 2024,
ApJ, 975, 292

\bibitem[Chequers et al.(2018)]{Chequers2018}
Chequers, M.~H., Widrow, L.~M., \& Darling, K. 2018,
MNRAS, 480, 4244

\bibitem[Chiba \& Sch{\"o}nrich(2021)]{Chiba2021}
Chiba, R. \& Sch{\"o}nrich, R. 2021,
MNRAS, 505, 2412

\bibitem[De Silva et al.(2015)]{DeSilva2015}
De Silva, G.~M., Freeman, K.~C., Bland-Hawthorn, J., Martell, S.,
de Boer, E.~W., Asplund, M., Keller, S., et al. 2015,
MNRAS, 449, 2604

\bibitem[Dehnen(2000)]{Dehnen2000}
Dehnen, W. 2000,
AJ, 119, 800

\bibitem[Frankel et al.(2023)]{Frankel2023}
Frankel, N., Bovy, J., Tremaine, S., \& Hogg, D.~W. 2023,
MNRAS, 521, 5917

\bibitem[Freeman \& Bland-Hawthorn(2002)]{Freeman2002}
Freeman, K. \& Bland-Hawthorn, J. 2002,
ARA\&A, 40, 487

\bibitem[Gaia Collaboration et al.(2018)]{GaiaCollaboration2018}
Gaia Collaboration, Katz, D., Antoja, T., Romero-G{\'o}mez, M., et al. 2018,
A\&A, 616, A11

\bibitem[Gaia Collaboration et al.(2023a)]{GaiaCollaboration2023}
Gaia Collaboration, Drimmel, R., Romero-G{\'o}mez, M., Chemin, L., et al. 2023a, A\&A, 674, A37

\bibitem[Gaia Collaboration et al.(2023c)]{Gaia2023Chem}
Gaia Collaboration, Recio-Blanco, A., Kordopatis, G.,
de Laverny, P., et al. 2023c,
A\&A, 674, A38

\bibitem[Gaia Collaboration et al.(2023b)]{GaiaDR3Summary2023}
Gaia Collaboration, Vallenari, A., Brown, A.~G.~A., Prusti, T.,
de Bruijne, J.~H.~J., Arenou, F., Babusiaux, C., et al. 2023b,
A\&A, 674, A1

\bibitem[Garavito-Camargo et al.(2019)]{GaravitoCamargo2019}
Garavito-Camargo, N., Besla, G., Laporte, C.~F.~P.,
Johnston, K.~V., G{\'o}mez, F.~A., \& Watkins, L.~L. 2019,
ApJ, 884, 51

\bibitem[Gilman et al.(2025)]{Gilman2025}
Gilman, D., Bovy, J., Frankel, N., \& Benson, A. 2025,
ApJ, 980, 24

\bibitem[G{\'o}mez et al.(2013)]{Gomez2013}
G{\'o}mez, F.~A., Minchev, I., O'Shea, B.~W., Beers, T.~C.,
Bullock, J.~S., \& Purcell, C.~W. 2013,
MNRAS, 429, 159

\bibitem[Helmi(2020)]{Helmi2020}
Helmi, A. 2020,
ARA\&A, 58, 205

\bibitem[Hunt et al.(2018)]{Hunt2018}
Hunt, J.~A.~S., Hong, J., Bovy, J., Kawata, D., \& Grand, R.~J.~J. 2018,
MNRAS, 481, 3794

\bibitem[Hunt et al.(2019)]{Hunt2019}
Hunt, J.~A.~S., Bub, M.~W., Bovy, J., Mackereth, J.~T.,
Trick, W.~H., \& Kawata, D. 2019,
MNRAS, 490, 1026

\bibitem[Hunt et al.(2020)]{Hunt2020}
Hunt, J. A. S., Johnston, K. V., Pettitt, A. R., et al. 2020,
MNRAS, 497, 818

\bibitem[Hunt et al.(2022)]{Hunt2022}
Hunt, J.~A.~S., Price-Whelan, A.~M., Johnston, K.~V., \& Darragh-Ford, E. 2022,
MNRAS, 516, L7

\bibitem[Hunt \& Vasiliev(2025)]{Hunt2025}
Hunt, J.~A.~S. \& Vasiliev, E. 2025,
New Astron. Rev., 100, 101721

\bibitem[Kawata et al.(2018)]{Kawata2018}
Kawata, D., Baba, J., Ciuc{\u a}, I., Cropper, M., Grand, R.~J.~J.,
Hunt, J.~A.~S., \& Seabroke, G. 2018,
MNRAS, 479, L108

\bibitem[Khachaturyants et al.(2022)]{Khachaturyants2022}
Khachaturyants, T., Beraldo e Silva, L., Debattista, V.~P.,
\& Daniel, K.~J. 2022,
MNRAS, 512, 3500

\bibitem[Khanna et al.(2019)]{Khanna2019}
Khanna, S., Sharma, S., Tepper-Garc{\'i}a, T., Bland-Hawthorn, J.,
Hayden, M.~R., Asplund, M., et al. 2019,
MNRAS, 489, 4962

\bibitem[Khoperskov et al.(2019)]{Khoperskov2019}
Khoperskov, S., Di Matteo, P., Gerhard, O., Katz, D.,
Haywood, M., Combes, F., Berczik, P., \& G{\'o}mez, A. 2019,
A\&A, 622, L6

\bibitem[Khoperskov et al.(2020)]{Khoperskov2020}
Khoperskov, S., Gerhard, O., Di Matteo, P., Haywood, M.,
Katz, D., Khrapov, S., Khoperskov, A., \& Arnaboldi, M. 2020,
A\&A, 634, L8

\bibitem[Khoperskov \& Gerhard(2022)]{Khoperskov2022}
Khoperskov, S. \& Gerhard, O. 2022,
A\&A, 663, A38

\bibitem[Laporte et al.(2018)]{Laporte2018}
Laporte, C.~F.~P., Johnston, K.~V., G{\'o}mez, F.~A.,
Garavito-Camargo, N., \& Besla, G. 2018,
MNRAS, 481, 286

\bibitem[Laporte et al.(2019)]{Laporte2019}
Laporte, C.~F.~P., Minchev, I., Johnston, K.~V., \& G{\'o}mez, F.~A. 2019,
MNRAS, 485, 3134


\bibitem[Lehmann et al.(2025)]{Lehmann2025} Lehmann, C., Murphy, M.~T., Liu, F.,
\& Flynn, C. 2025,
MNRAS, 536, 498

\bibitem[Li et al.(2025)]{Li2025}
Li, Y., Freeman, K., Jerjen, H., Buder, S., Hayden, M., \& Mondal, A. 2025,
MNRAS, 538, 1963

\bibitem[Lindegren et al.(2021)]{Lindegren2021}
Lindegren, L., Klioner, S.~A., Hern{\'a}ndez, J., Bombrun, A.,
Ramos-Lerate, M., Steidelm{\"u}ller, H., Bastian, U., et al. 2021,
A\&A, 649, A2

\bibitem[Martell et al.(2017)]{Martell2017}
Martell, S.~L., Sharma, S., Buder, S., Duong, L., Schlesinger, K.~J.,
Simpson, J.~D., Lind, K., et al. 2017,
MNRAS, 465, 3203

\bibitem[McMillan(2017)]{McMillan2017}
McMillan, P.~J. 2017,
MNRAS, 465, 76

\bibitem[Minchev \& Famaey(2010)]{MinchevFamaey2010}
Minchev, I. \& Famaey, B. 2010,
ApJ, 722, 112

\bibitem[Monari et al.(2019)]{Monari2019}
Monari, G., Famaey, B., Siebert, A., et al. 2019,
A\&A, 632, A107

\bibitem[Ramos et al.(2018)]{Ramos2018}
Ramos, P., Antoja, T., \& Figueras, F. 2018,
A\&A, 619, A72

\bibitem[Rix \& Bovy(2013)]{Rix2013}
Rix, H.-W. \& Bovy, J. 2013,
A\&ARv, 21, 61

\bibitem[Sanders \& Binney(2016)]{SandersBinney2016}
Sanders, J.~L. \& Binney, J. 2016,
MNRAS, 457, 2107

\bibitem[Sellwood \& Binney(2002)]{SellwoodBinney2002}
Sellwood, J.~A. \& Binney, J.~J. 2002,
MNRAS, 336, 785

\bibitem[Sharma et al.(2018)]{Sharma2018}
Sharma, S., Stello, D., Bland-Hawthorn, J., Huber, D.,
Bedding, T.~R., et al. 2018,
MNRAS, 473, 2004

\bibitem[Sheinis et al.(2015)]{Sheinis2015}
Sheinis, A., Anguiano, B., Asplund, M., Bacigalupo, C.,
Barden, S., Birchall, M., Bland-Hawthorn, J., et al. 2015,
J. Astron. Telesc. Instrum. Syst., 1, 035002

\bibitem[Tepper-Garc{\'i}a et al.(2025)]{TepperGarcia2025}
Tepper-Garc{\'i}a, T., Bland-Hawthorn, J.,
Bedding, T.~R., Federrath, C., \& Agertz, O. 2025,
MNRAS, 542, 1987

\bibitem[Trick et al.(2019)]{Trick2019}
Trick, W.~H., Coronado, J., \& Rix, H.-W. 2019,
MNRAS, 484, 3291

\bibitem[Trick(2022)]{Trick2022}
Trick, W.~H. 2022,
MNRAS, 509, 844

\bibitem[Tremaine et al.(2023)]{Tremaine2023}
Tremaine, S., Frankel, N., \& Bovy, J. 2023,
MNRAS, 521, 114

\bibitem[Vasiliev(2019)]{Vasiliev2019}
Vasiliev, E. 2019,
MNRAS, 482, 1525

\bibitem[Wang et al.(2020)]{Wang2020}
Wang, H.-F., Huang, Y., Zhang, H.-W., et al. 2020,
ApJ, 902, 70

\bibitem[Wang et al.(2026)]{Wang2026}
Wang, S., Brown, A.~G.~A., Debattista, V.~P., \& Khachaturyants, T. 2026,
A\&A, 710, A120

\bibitem[Xu et al.(2026)]{Xu2026}
Xu, Y., Liu, C., Li, C. D., et al. 2026,
ApJ, 1006, 143,
doi:10.3847/1538-4357/ae7f0e

\bibitem[Yamsiri et al.(2026)]{Yamsiri2026}
Yamsiri, P., Bland-Hawthorn, J., \& Tepper-Garc{\'i}a, T. 2026,
arXiv e-prints, arXiv:2602.05296

\bibitem[Yang et al.(2023)]{Yang2023}
Yang, P., Wang, H.-F., Luo, Z.-Q., et al. 2023,
AJ, 165, 110

\end{thebibliography}
\end{document}